\ifx\onecol\undefined
\documentclass[10pt,journal,twocolumn]{IEEEtran}
\else 
\documentclass[12pt,draftclsnofoot,journal,onecolumn]{IEEEtran}
\fi

\IEEEoverridecommandlockouts

\usepackage{amsmath}
\usepackage{amsthm,amssymb}
\usepackage{algorithm}
\usepackage{algorithmic}
\usepackage{bm,bbm}
\usepackage{balance}
\usepackage{braket}
\usepackage{booktabs}
\usepackage{color}
\usepackage{cases}
\usepackage{eqparbox}
\usepackage{enumitem}
\usepackage{graphicx}
\usepackage{mathrsfs}
\usepackage{multirow}
\usepackage{subfigure}
\usepackage{setspace}
\usepackage{threeparttable}
\usepackage{verbatim}

\usepackage[numbers,sort&compress]{natbib}

\newtheorem{lemma}{Lemma}
\newtheorem{theorem}{Theorem}
\newtheorem{remark}{Remark}

\newtheorem{definition}{Definition}
\newtheorem{proposition}{Proposition}

\usepackage{hyperref}
\hypersetup{hidelinks, 
colorlinks=true,
allcolors=black,
pdfstartview=Fit,
breaklinks=true}

\newcommand{\ri}{{\mathrm{i}}}

\DeclareMathOperator*{\argmax}{arg\,max}

\def \T {^{\mathsf{T}}}
\def \H {^{\mathsf{H}}}
\def \sinc {{\rm sinc}}

\begin{document}

\title{MIMO Capacity Analysis and Channel Estimation for Electromagnetic Information Theory}
% \title{ Ergodic Capacity Saturation Predicted by Electromagnetic Information Theory:\\ Limitation or Opportunity? }

\author{{Jieao~Zhu,~Vincent~Y.~F.~Tan,~{\textit{Senior~Member,~IEEE}},~and~Linglong~Dai,~{\textit{Fellow,~IEEE}}\vspace{-0.75cm} }
\thanks{This work was supported in part by the National Natural Science Foundation of China (Grant No. 62325106) and in part by the National Key Research and Development Program of China (Grant No.2023YFB3811503). {\em (Corresponding author: Linglong Dai.)}}
\thanks{J. Zhu and L. Dai are with the Department of Electronic Engineering, Tsinghua University, Beijing 100084, China and also with the Beijing National Research Center for Information Science and Technology (BNRist), Beijing 100084, China (e-mails: zja21@mails.tsinghua.edu.cn, daill@tsinghua.edu.cn).}
\thanks{Vincent Y. F. Tan is with the Department of Mathematics and the Department of Electrical and Computer Engineering, National University of Singapore, Singapore 119077, Singapore (e-mail: vtan@nus.edu.sg). }
}

% \markboth{To be Submitted to IEEE Transactions on Wireless Communications}{}

\maketitle

\begin{abstract}
    Electromagnetic information theory (EIT) is an interdisciplinary subject that serves to integrate deterministic electromagnetic theory with stochastic Shannon's information theory. Existing EIT analysis operates in the continuous space domain, which is not aligned with the practical algorithms working in the discrete space domain. This mismatch leads to a significant difficulty in application of EIT methodologies to practical discrete space systems, which is called as the {\em discrete-continuous gap} in this paper. To bridge this gap, we establish the discrete-continuous correspondence with a prolate spheroidal wave function (PSWF)-based ergodic capacity analysis framework. Specifically, we state and prove some discrete-continuous correspondence lemmas to establish a firm theoretical connection between discrete information-theoretic quantities to their continuous counterparts. With these lemmas, we apply the PSWF ergodic capacity bound to advanced MIMO architectures such as continuous-aperture MIMO (CAP-MIMO) and extremely large-scale MIMO (XL-MIMO). From this PSWF capacity bound, we discover the capacity saturation phenomenon both theoretically and empirically. Although the growth of MIMO performance is fundamentally limited in this EIT-based analysis framework, we reveal new opportunities in MIMO channel estimation by exploiting the EIT knowledge about the channel. Inspired by the PSWF capacity bound, we utilize continuous PSWFs to improve the pilot design of discrete MIMO channel estimators, which is called as the PSWF channel estimator (PSWF-CE). Simulation results demonstrate improved performances of the proposed PSWF-CE, compared to traditional minimum mean squared error (MMSE) and compressed sensing-based estimators. 
\end{abstract}
    %\vspace{-2em}

\section{Introduction}
Future sixth-generation (6G) wireless communication systems are expected to form the backbone of various emerging applications, such as digital twins, autonomous driving, and the internet-of-things. To meet the high spectral efficiency requirement of 6G, it is necessary to fully unleash the information-carrying capability of the electromagnetic (EM) fields via advanced multi-input multi-output (MIMO) technologies. % For example, massive MIMO (mMIMO) and extra-large MIMO (XL-MIMO) technologies are devoted to enlarging the array aperture for improved spatial multiplexing gain. In addition to array aperture expansion, the holographic MIMO (H-MIMO) technology is expected to fully exploit the spatial DoF by densifying the MIMO aperture. To further shed light on the development of MIMO communications, many researchers are trying to rethink the fundamentals that underlie all the wireless communication processes, which motivates the study of electromagnetic information theory (EIT). 
On one hand, electromagnetics, which is crowned by the well-known Maxwell's equations, governs all EM phenomena at the macroscopic level in which wireless communication systems operate. On the other hand, information theory, invented by Claude E.\ Shannon, reveals in a probabilistic way the exact rate boundary under which error-free information transmission is possible. 
Although both electromagnetics and information theory significantly contribute to design and optimization of modern wireless systems, these two parts of solid knowledge are loosely coupled in most of the existing literature~\cite{migliore2018horse}. 
This motivates the study of {\em  electromagnetic information theory} (EIT), an emerging interdisciplinary subject aiming to reveal the amount of information that EM fields are capable of carrying~\cite{zhu2024electromagnetic}.

\subsection{Prior Works}
% Theoretical analysis
Existing works on EIT can be generally divided into two categories: Theoretical EIT  analysis, and the applications of EIT to practical systems. 

{\bf Theoretical  EIT Analysis.}
The attempt to establish a unified electromagnetics-based information transmission theory dates back to 1946~\cite{gabor1946theory} even in the pre-Shannon era. Early works on EIT analysis  concentrated on the interpolation and extrapolation of scattered EM fields~\cite{bucci1987spatial,bucci1989degrees,migliore2008electromagnetics,li2022degrees}, in which the optimal number of basis functions and the minimum number of parameters required to approximate a scattered EM field was established. These numbers are usually referred to as the {\em electromagnetic degrees-of-freedom} (EM-DoF) of the system under investigation, and is usually derived by upper bounding the difference between the EM Green's operator and its lowpass-filtered version through an asymptotic spatial bandwidth analysis~\cite{ding2022degrees,bucci1987spatial,bucci1989degrees,li2022degrees}. These EM-DoF results usually hold in the asymptotic regime, where the volume of the observation region (receiver) tends to infinity.  
In contrast to the asymptotic EM-DoF analysis, another salient information-theoretic quantity is the effective DoF (eDoF) derived from the functional approximation theory~\cite{franceschetti2017wave,ruiz2023degrees}. The eDoF is defined as the number of significant eigen-modes of some EM transmission operator above a prescribed threshold~\cite{dardari2020communicating,ruiz2023degrees}. The eDoF analysis of paraxial line-of-sight (LoS) channels is conducted in~\cite{miller2000communicating}, where the prolate spheroidal wave function (PSWF) approach was adopted to diagonalize the transmission operator. The PSWF-based method was extended to the general non-paraxial LoS case in~\cite{ruiz2023degrees}. To further capture the non-line-of-sight (NLoS) EM propagation, the authors of~\cite{poon2005degrees} established an EM-based deterministic operator channel model, and derived the first-order maximum DoF of this channel. Interestingly, this derivation was also deeply rooted in PSWFs. The capacity of the same channel was derived in~\cite{jensen2008capacity}. To relax the assumption of deterministic channels, the authors of~\cite{poon2006impact} proposed the first stochastic EM channel model, where the channel was characterized by a white Gaussian random field in the wavenumber domain. An additional bandlimited prior is assumed to this stochastic channel in~\cite{nam2014capacity}, where the ergodic capacity upper bound was derived from the bandlimited model via PSWFs.

{\bf EIT Applications.} 
Recently, the concept of EIT has caught the attention of many researchers due mainly to the need for a unified analysis framework for   various beyond massive MIMO technologies for 6G~\cite{bjornson2024towards}, including continuous-aperture MIMO (CAP-MIMO)~\cite{wei2023tri,sanguinetti2022wavenumber}, extremely large-scale MIMO (XL-MIMO)~\cite{lu2023near}, fluid antenna systems~\cite{wong2023fluid}, superdirective antenna arrays~\cite{marzetta2019super}, and reconfigurable intelligent surfaces~\cite{liu2021reconfigurable}. To accurately characterize the electromagnetic propagation channel, electromagnetic-compliant channel modeling~\cite{wang2022electromagnetic} has become a mainstream research direction of EIT channel modeling.
The Green's function-based deterministic LoS channel has been widely accepted due to its simplicity~\cite{dardari2020communicating}, where stochastic fading properties of the practical channels are neglected in the modeling. Borrowing mathematical techniques for the stochastic channel models~\cite{poon2006impact,shafi2006polarized}, electromagnetic random field channel modeling was recently explored in~\cite{pizzo2020spatially,pizzo2022spatial}. Extensions to the two-dimensional wavenumber domain~\cite{pizzo2022spatial,pizzo2022fourier} and applications to capacity computation of wideband polarized electromagnetic channels~\cite{mikki2023shannon} were discussed. To characterize the complex rich-scattering electromagnetic environment, the authors of~\cite{li2023electromagnetic} proposed a group-$T$-matrix based electromagnetic channel model with cylindrical harmonics. 
Based on the established electromagnetic-compliant channel models, further efforts have been devoted to beamforming~\cite{wei2023tri,sanguinetti2022wavenumber,castellanos2023electromagnetic}, precoding designs~\cite{wei2022multi}, and channel estimation~\cite{ghermezcheshmeh2023parametric} of practical systems.

% The following two sentences are too specific here. 
% Following the EM-compliant channel model in~\cite{pizzo2022fourier}, the authors of~\cite{wei2022multi} designed several downlink multi-user MIMO precoding algorithms. 
% Furthermore, in~\cite{castellanos2023electromagnetic}, a high-gain beamforming method with Hertzian dipole-based antenna models was proposed, taking into consideration the near-field propagation. To incorporate the tri-polarization characteristics, a two-layer precoding algorithm for tri-polarized multi-user H-MIMO systems was proposed~\cite{wei2023tri}. 

% In Shannon's seminal paper~\cite{shannon1948mathematical}, he defined the modern concept of information via a statistical approach, and established three fundamental theorems regarding the theoretical limits of lossless source coding, lossy source coding, and noisy channel coding. It is Shannon's work that have enabled wireless communications, since without channel codes it is almost impossible to convey long sequences of information to the receiver. 
% Thanks to the invention of various modulation techniques~\cite{pasupathy1979minimum}, the time-frequency wireless channel is converted to a sequence of nearly independent channel uses, enabling the application of channel coding. 

% Method: You should find out the papers that you have read. 
% Gather them into a folder. Try to find out the underlying logic thread. 
% Then try to write this introduction part. 

Based on the above summary of the state-of-the-art in the field, we can observe that the theoretical and practical algorithmic aspects of EIT have generally developed in parallel. On one hand, the theoretical DoF~\cite{bucci1989degrees,poon2005degrees} and capacity results~\cite{gruber2008new,nam2014capacity,dardari2020communicating} are only consistent in the continuous system regime. Although we may intuitively expect it to be true, it is not {\em a priori} clear whether practical discrete MIMO systems can approach such a continuous theoretical limit. It is also not clear under what conditions these bounds for continuous systems can be directly applied to practical discrete systems. On the other hand, most of the EIT-related applications~\cite{wei2023tri,sanguinetti2022wavenumber} adopt direct discretization of the electromagnetically compliant integral equation-based models. Although the integrals can generally be approximated by  sums, the information-theoretic inequalities will possibly be violated as a result of this discrete approximation procedure. This issue is more pronounced when operating in the asymptotic regimes where ultra-dense antennas (also known as CAP-MIMO and H-MIMO) or extremely large array apertures (XL-MIMO) are considered. 
The root cause of these inconsistencies lie in the absence of a unified comparison framework between discrete matrix-based MIMO systems and continuous operator-based systems, which we term as the {\it discrete-continuous gap}.

\subsection{Our Contributions}
To fill in this gap, in this paper, we prove several {\it discrete-continuous correspondence lemmas} with a PSWF-based ergodic capacity analysis framework, thus enabling the application of the PSWF bounds to practical H-MIMO and XL-MIMO systems. Furthermore, we uncover new opportunities for utilizing continuous PSWFs to improve discrete channel estimators.\footnote{Simulation codes to reproduce the numerical results in this paper will be provided at the following link: \url{http://oa.ee.tsinghua.edu.cn/dailinglong/publications/publications.html}} The contributions of this paper are summarized as follows. 

\begin{itemize}[leftmargin=10pt]
    \item By establishing the continuous transmission model of EIT, we prove discrete-continuous correspondence lemmas. Specifically, the $\varepsilon$-DoF and ergodic capacity of an arbitrary discrete MIMO system are dominated by those of a continuous MIMO system of the same power budget and noise level. This theoretical correspondence lemma establishes a natural comparison between discrete and continuous systems, thus allowing existing continuous theoretical methodologies to be applied to practical discrete MIMO systems. 
    \item Based on the discrete-continuous correspondence lemmas, we further prove that the ergodic capacity of both H-MIMO and XL-MIMO systems will saturate to some unified theoretical upper limit under an EM-compliant wavenumber domain correlated channel model. We numerically evaluate the unified upper limit, which we term the {\em EIT-PSWF bound}, by using the spectral method and the technique of perturbation to PSWF integral equations. 
    \item Although the EIT-PSWF bound limits the growth of the MIMO ergodic capacity, inspired by the discrete-continuous correspondence, we propose a PSWF-based MIMO channel estimator (PSWF-CE) with improved performance. Different from traditional compressed sensing (CS) estimators where the pilots are randomly generated, the proposed estimator adopts the PSWF waveforms as  pilot sequences, which exploits the bandlimited structure of the electromagnetic channel. Simulation results show the improved performance of the proposed PSWF-CE compared to traditional minimum mean-squared error (MMSE) estimators and CS-based estimators.  
\end{itemize}

\subsection{Organization and Notation}
\emph{Organization}:
The rest of the paper is organized as follows.
Preliminaries on electromagnetics and PSWFs are introduced in Section~\ref{sec1}. System transmission models, wavenumber domain correlated channel models, and the problem of MIMO channel estimation are introduced in Section~\ref{sec3-System-Models-and-Problem-Formulation}. The main theoretical results concerning the discrete-continuous correspondence and the capacity saturation are presented in Section~\ref{sec4}. PSWF evaluation techniques and the numerical saturation results are provided in Section~\ref{sec5}. The proposed PSWF-CE MIMO channel estimator is introduced in Section~\ref{sec6}. Numerical results are reported in Section~\ref{sec7}. Finally, Section~\ref{sec8} concludes this paper and suggests directions for future research. 

\emph{Notation}: 
The symbol $[N]$ denotes the set of integers  $\{1,2,\cdots, N\}$; 
bold uppercase characters ${\bf X}$ denote matrices, with $[{\bf X}]_{mn}$ representing its $(m,n)$--th entry; 
bold lowercase characters ${\bf x}$ denote vectors;
${\bf I}_n$ denotes the identity matrix of size $n$; 
${\bf X}\H, {\bf X}\T$, and ${\bf X}^*$ denotes Hermitian transpose, transpose, and complex conjugate of ${\bf X}$, respectively; 
for two operators $\mathcal{L}_1$ and $\mathcal{L}_2$, $\{\mathcal{L}_1, \mathcal{L}_2\}$ denotes the commutator $\mathcal{L}_1\mathcal{L}_2 - \mathcal{L}_2\mathcal{L}_1$; 
$\|\cdot\|$ denotes the 2-norm of a matrix, an operator, or the $\mathcal{L}^2$--norm of a function; 
$\|\cdot\|_{\rm F}$ denotes the Frobenius norm of a matrix; 
$|\cdot|$ denotes the cardinality of finite sets or the length of its complex argument; 
${\rm vec}({\bf X})$ stacks the columns of the matrix $\bf X$ into a single column vector; 
$\mathcal{L}^2(K)$ denotes the set of  Lebesgue square integrable functions supported on $K$; 
$\ell^2$ denotes all the square summable sequences; 
$\dagger$ denotes the adjoint operator; 
$\sigma_i({\bf H})$ and $\sigma_i(T)$ denote the $i$-th ($i\geq 0$) largest singular value of matrix ${\bf H}$ and compact linear operator $T$, respectively; 
$\circ$ denotes operator composition; 
$\mathcal{CN}({\bm \mu}, {\bf \Sigma})$ denotes the circularly Gaussian distribution with mean ${\bm \mu}$ and covariance ${\bf \Sigma}$;  and finally,
${\rm supp}(f) = \{x\in\mathcal{X}:f(x)\ne 0\}$ denotes the support of a function $f:\mathcal{X}\to\mathbb{R}$.

\section{Preliminaries} \label{sec1}
In this section, we introduce the electromagnetic background knowledge for EIT, with a special focus on the continuous space electromagnetic propagation laws and the analysis of bandlimited functions. 

\subsection{Electromagnetic Propagation and Electromagnetic Channels}
All wireless channel characteristics stem from the underlying electromagnetic propagation laws in the form of Maxwell's four partial differential equations (PDEs)~\cite{kong1975theory}. Following Maxwell's procedure, by operating in the Fourier transform domain with the $e^{-\ri \omega t}$ convention, we arrive at the frequency-domain vector wave equation 
\begin{equation}
    \nabla\times\nabla\times {\bf E} - \omega^2 \mu\epsilon {\bf E} = \ri \omega\mu {\bf J}_f, \label{eqn:VWE_FreqDomain}
\end{equation}
where $\omega$ is the carrier frequency, $\bf E$ is the received electric field, ${\bf J}_f$ is the transmitted free current density, $\mu$ is the magnetic permeability, and $\epsilon$ is the dielectric permittivity. The general solution to the linear vector PDE~\eqref{eqn:VWE_FreqDomain} can be obtained by the {\it method of Green's function}, in which the solution ${\bf E}({\bf x},\omega)$ is of the form 
\begin{equation}
    {\bf E}({\bf x},\omega) = \ri\omega\mu \iiint_{V_t} {\bf G}({\bf x, x'};\omega) {\bf J}_f ({\bf x'},\omega) {\rm d}^3 {\bf x'}, \quad {\bf x}\in V_r\label{eqn:from_J_to_E}
\end{equation} 
where $V_t\subset \mathbb{R}^3$ is the region in which the source lies, $V_r$ is the region where the receiver lies, and ${\bf G}\in\mathbb{C}^{3\times 3}$ is the dyadic Green's function expressed in Cartesian coordinates as 
\begin{equation}
    {\bf G}({\bf x}, {\bf x'}) = \left({\bf I}_3 + \frac{1}{k_c^2} \nabla\nabla \right)g(\|{\bf x-x'}\|). \label{eqn:GreenFunction}
\end{equation}
The scalar $k_c = \omega\sqrt{\mu\epsilon}=2\pi/\lambda_c$ is called the free-space wavenumber, $\lambda_c$ is the carrier wavelength, and $g(r)$ is the scalar Green's function expressed as $g(r) = \frac{e^{\ri k_cr}}{4\pi r}$, 
where $r = \|{\bf x-x'}\|$. An important property of the scalar Green's function is given by Weyl's identity~\cite{pizzo2022spatial,pizzo2020spatially}
\begin{equation}
    g(r) = \frac{\ri}{8\pi^2}\int_{-\infty}^{\infty}{\rm d}k_x \int_{-\infty}^{\infty} {\rm d}k_y \frac{e^{\ri (k_x x + k_y y + k_z |z|)}}{k_z}, \label{eqn:Weyl}
\end{equation} 
where $k_z = \sqrt{k_c^2 - k_x^2 -k_y^2}$ for $k_x^2 + k_y^2 < k_c^2$ and $k_z = \ri \sqrt{k_x^2 + k_y^2 - k_c^2}$ for $k_x^2 + k_y^2 > k_c^2$. 

\begin{remark} \label{remark:Electromagnetics}
    The electromagnetic signal propagation law in isotropic media
    is fully characterized by~\eqref{eqn:from_J_to_E}, where the dyadic Green's function ${\bf G}$ is intrinsically determined by the scalar Green's function $g$, whose integral representation is given by Weyl's identity~\eqref{eqn:Weyl}. An important observation from~\eqref{eqn:Weyl} is that, as the electromagnetic wave propagates along the $z$-axis, the observed field will tend to a $k_c$-bandlimited one~\cite{pizzo2022fourier}, i.e., 
    \begin{equation}
        \begin{aligned}
            {\rm supp}\{\mathcal{F}^{(2)}[g]({\bf k}^{(2)})\} \to & \, \{{\bf k}^{(2)} = (k_x, k_y): k_x^2 + k_y^2 < k_c^2\}, \\
            & \quad{\rm as}\, |z|\to\infty, 
        \end{aligned}
    \end{equation}
    where the limit is interpreted in the sense of energy concentration, and the above Fourier transform $\mathcal{F}^{(2)}[\cdot]$ is taken w.r.t. coordinates $(x,y)$ only. 
\end{remark}

\subsection{Prolate Spheroidal Wave Functions (PSWF)}
As is pointed out in {\bf Remark~\ref{remark:Electromagnetics}}, wave propagation phenomena are rich sources of bandlimited functions~\cite{pizzo2022spatial,franceschetti2017wave}. A bandlimited function $f$ is a square integrable function whose Fourier transform is supported on a bounded set. It is well-known that $f$ and its Fourier transform $\tilde{f}$ cannot be simultaneously concentrated both in the time and the frequency domain~\cite{slepian1964prolate4}. Thus, we need to find the ``most concentrated'' functions that best satisfy time-frequency concentration. The task of finding these most concentrated functions is termed {\it Slepian's concentration problem}~\cite{slepian1964prolate4}, and it turns out that the solution to this problem is the prolate spheroidal wave functions (PSWFs). 

Before introducing the PSWFs, we need to first introduce an important theorem that ensures the existence of such functions. This theorem is widely applicable to continuous, discrete, and even random concentration problems.  

\begin{theorem}[Simultaneous orthogonalization~\cite{ihara1993information}] \label{thm:SimultaneousOrthogonalization}
    Let $\mathcal{H}$ be a Hilbert space. Let  $\mathcal{H}_1$ and $\mathcal{H}_2$ be two subspaces of $\mathcal{H}$, and let $\Pi_1, \Pi_2$ be two projection operators that project every element $x\in\mathcal{H}$ to $\Pi_1 x\in \mathcal{H}_1$ and $\Pi_2 x\in \mathcal{H}_2$. Let $A = \Pi_1 \Pi_2 \Pi_1$ and $B = \Pi_2 \Pi_1 \Pi_2$. Then, there exist complete orthonormal bases $\{\phi_n\}_{n=0}^\infty$ of $\mathcal{H}_1$ and $\{\psi_n\}_{n=0}^\infty$ of $\mathcal{H}_2$ that span the eigenspaces of the operators $A$ and $B$ respectively, and the eigenvalues of $A,B$ are given by the same sequence of non-negative real numbers $\{\lambda_n\}_{n=0}^\infty\subset [0,1]$. In particular, $\{\phi_n\}_{n=0}^\infty$, $\{\psi_n\}_{n=0}^\infty$, and $\{\lambda_n\}_{n=0}^\infty$ satisfy:
    \begin{equation}
        \begin{aligned}
            A\phi_n &= \lambda_n \phi_n,  \\
            B\psi_n &= \lambda_n \psi_n,  \\
            \langle \phi_m, \psi_n\rangle &= \sqrt{\lambda_n}\delta_{mn}, \quad \forall n,m=0,1,2.\cdots. 
        \end{aligned}
    \end{equation}
\end{theorem}

\begin{remark}
     {\bf Theorem~\ref{thm:SimultaneousOrthogonalization}} implies that $\Pi_2 \phi_n = \sqrt{\lambda_n} \psi_n$. Furthermore, if the eigenvalues are non-degenerate, then the choices of $\{\phi_n\}_{n=0}^\infty$ and $\{\psi_n\}_{n=0}^\infty$ are unique (up to a global phase). 
\end{remark}

With this simultaneous orthogonalization theorem, we can define the $k$-dimensional PSWFs. Specifically, let $(\mathcal{A}, \mathcal{B})$ be a pair of compact sets in $\mathbb{R}^k$. Let the Hilbert space $\mathcal{H}_1$ be spanned by all square-integrable functions supported on $\mathcal{A}$, i.e., 
\begin{equation}
    \mathcal{H}_1 := \{f\in \mathcal{L}^2(\mathbb{R}^k): {\rm supp}(f) \subset \mathcal{A}\}. 
    \label{eqn:constraint_f} 
\end{equation} 
Similarly, let the Hilbert space $\mathcal{H}_2$ be spanned by all square-integrable functions whose Fourier transform are supported on $\mathcal{B}$, i.e., 
\begin{equation}
    \mathcal{H}_2 := \{g\in\mathcal{L}^2(\mathbb{R}^k): {\rm supp}(\mathcal{F}[g])\subset \mathcal{B}\}.  
    \label{eqn:constraint_g}
\end{equation} 
Then we can define the PSWFs as follows. 
\begin{definition}[$k$-dimensional PSWFs]
    A sequence of $k$-dimensional $(\mathcal{A}, \mathcal{B})$-PSWFs $(\psi_n, \lambda_n)_{n=0}^{\infty}$ consists the sequence of eigenfunctions $\{\psi_n\}_{n=0}^\infty$ and the sequence of eigenvalues $\{\lambda_n\}_{n=0}^\infty$, where both $\{\psi_n\}_{n=0}^{\infty}$ and $\{\lambda_n\}_{n=0}^\infty$ are constructed from {\bf Theorem~\ref{thm:SimultaneousOrthogonalization}} by instantiating $\mathcal{H} = \mathcal{L}^2(\mathbb{R}^k)$ and $\mathcal{H}_1, \mathcal{H}_2$ as in~\eqref{eqn:constraint_f} and~\eqref{eqn:constraint_g}. 
\end{definition}
Of practical importance is the one-dimensional case $k=1$, where many useful properties of 1D-PSWFs can be derived. In the case $k=1$, the set $\mathcal{A}$ is usually a finite union of intervals, and the set $\mathcal{B}$ is usually an interval centered at 0. In the following, we write $\mathcal{B}_W$ for the centered symmetric interval $[-W/2, W/2]$ of bandwidth $W>0$. 
\begin{theorem} [Properties of 1D-PSWFs~\cite{nam2014capacity}] \label{thm:Properties_1D_PSWFs}
    One-dimensional $(\mathcal{A}, {\mathcal{B}_W})$-PSWFs $(\psi_n, \lambda_n)_{n=0}^\infty$ satisfy the following four properties: 

    1) The functions $\psi_n$ are real and bandlimited in $\mathcal{B}_W$. 

    2) The functions $\psi_n$ are orthonormal on $\mathbb{R}$ and complete in the space of $\mathcal{H}_2$ containing all $\mathcal{B}_W$-bandlimited functions. 

    3) The functions $\psi_n$ are orthogonal on $\mathcal{A}$ and complete in $\mathcal{L}^2(\mathcal{A})$. Specifically, 
    \begin{equation}
        \int_{\mathcal{A}} \psi_{m}(x)\psi_{n}(x) {\rm d}x = \lambda_m \delta_{mn}. 
    \end{equation}

    4) The low-pass filtered version of the PSWF satisfies the following eigen-integral equation 
    \begin{equation}
        \int_{\mathcal{A}} W \sinc (W(x-y)) \psi_n(y) {\rm d}y = \lambda_n \psi_n(x), \quad\forall\, x\in\mathbb{R}, \label{eqn:PSWF_integral_equation}
    \end{equation}
    where $\sinc (x) := \sin(\pi x)/(\pi x)$. 
\end{theorem}
The proof of {\bf Theorem~\ref{thm:Properties_1D_PSWFs}} is a direct application of the simultaneous orthogonalization {\bf Theorem~\ref{thm:SimultaneousOrthogonalization}} to the definition of 1D-PSWFs.  

\begin{figure*}[t]
    \centering
    \includegraphics[width=0.96\linewidth]{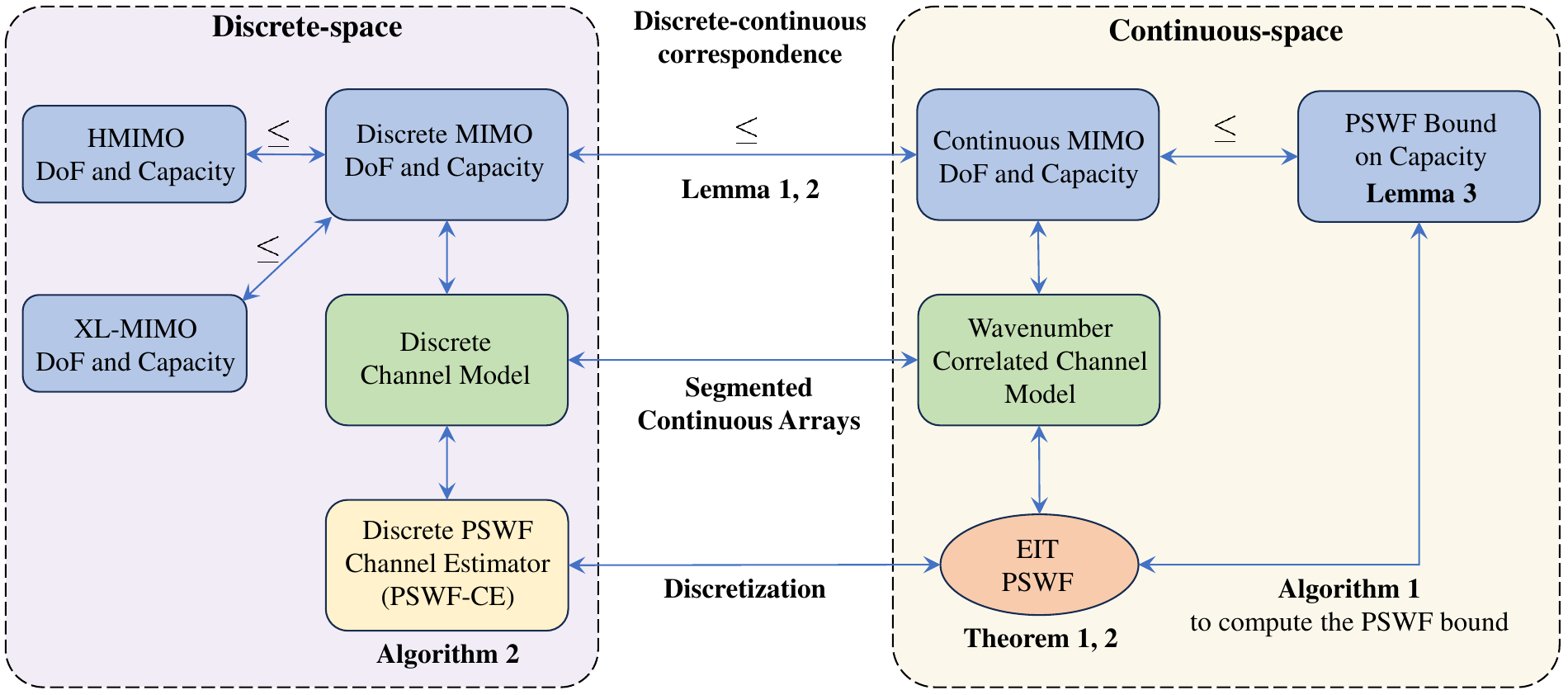}
    \caption{Organization of this paper. }
    \label{fig:Concepts}
\end{figure*}
% After revision, I will try to convert this pdf figure to tex codes. 
% I will continuously update the .pdf files according to your suggestions. 

\section{System Models and Problem Formulation} \label{sec3-System-Models-and-Problem-Formulation}
In this section, we will first specify both the discrete space and the continuous space transmission models in wireless communications. Then, we will elaborate on the continuous space channel model with bandlimited property in the wavenumber domain.  

\subsection{Discrete and Continuous Signal Models}
Let us consider a narrowband time-invariant fully digital discrete MIMO transmission model with $N_t$ transmitter (Tx) antennas and $N_r$ receiver (Rx) antennas. The input-output relationship between the transmitted vector ${\bf x}\in\mathbb{C}^{N_t}$ and the received vector ${\bf y}\in\mathbb{C}^{N_r}$ is given by 
\begin{equation}
    {\bf y} = {\bf Hx} + {\bf z}, \label{eqn:DMIMO_channel}
\end{equation}
where ${\bf H}\in\mathbb{C}^{{N_t}\times N_r}$ is the channel matrix, and ${\bf z}\in\mathbb{C}^{N_r}$ is an additive white Gaussian noise (AWGN) random vector with distribution ${\mathcal{CN}}({\bf 0}, \sigma_z^2 {\bf I}_{N_r})$. The vector ${\bf x}$ is subject to a power constraint $\mathbb{E}[{\bf x}\H {\bf x}]\leq P_{\rm T}$. For any fixed $\bf H$, the capacity of the discrete MIMO model~\eqref{eqn:DMIMO_channel} is given by 
\begin{equation}
    C_{\rm dMIMO} = \max_{{\bf C_x}\succeq 0: {\rm tr}({\bf C_x})\leq P_{\rm T}}\log\det\bigg({\bf I} + \frac{1}{\sigma_z^2}{\bf HC_x H}\H \bigg), \label{eqn:MIMO_Capacity}
\end{equation}
and the capacity can be achieved by the well-known singular value decomposition (SVD)-based precoding/combining. 

In contrast to the above discrete MIMO model, real-world MIMO transceivers operate on continuous electromagnetic fields~\cite{sanguinetti2022wavenumber}. 
Thus, we consider a single-user continuous MIMO communication scenario where a multi-antenna base station (BS) serves a multi-antenna user. To be general, we do not assume any specific antenna structure (H-MIMO, XL-MIMO, discrete MIMO, etc.) at either the BS or the user. Inspired by~\eqref{eqn:from_J_to_E}, we commence by establishing a continuous electromagnetic level signal transmission model between the source and the destination, which is given by the following continuous space linear array signal model~\cite{poon2005degrees,poon2006impact}
\begin{equation}
    Y(q) = \int_{{\bf L}_t} h(q,p) X(p) {\rm d}p +Z(q), \quad q\in {\bf L}_r, \label{eqn:H-MIMO_signalModel}
\end{equation}
where ${X}\in \mathcal{L}^2({\bf L}_t)$ is the transmitted field, $Y\in\mathcal{L}^2({\bf L}_r)$ and $Z\in\mathcal{L}^2({\bf L}_r)$ are the received field and the white noise field with autocorrelation $\sigma_z^2\delta(q-q')$, respectively, and $h\in \mathcal{L}^2({\bf L}_r\times {\bf L}_t)$ is the continuous space channel that links Tx to Rx, and ${\bf L}_t, {\bf L}_r\subset \mathbb{R}$ are one-dimensional transmitter and receiver intervals, respectively. Note that in this model, we avoid using the complicated electromagnetic vector-valued representation~\eqref{eqn:GreenFunction} to facilitate further analysis. Here the notations are aligned to~\cite{nam2014capacity} for clarity. Accurate electromagnetic models of the propagation channels with random scatterers~\cite{bucci1987spatial}, antenna mutual coupling, and fading effects are practically meaningful. However, in this paper, we aim to identify the most significant factors that contribute to the fundamental limits of MIMO systems, which is determined by the bandlimited behavior of the channel $h$. Since vector-valued polarization and mutual coupling only contribute to at most a constant factor in the capacity formula, we ignore these effects and focus on a simpler case of a monotone scalar field model~\eqref{eqn:H-MIMO_signalModel}.

For such a continuous transmission model~\eqref{eqn:H-MIMO_signalModel} with transmit power constraint $\mathbb{E}[\int_{{\bf L}_t}|X(p)|^2 {\rm d}p]\leq P_{\rm T}$, the capacity is given by~\cite{ihara1993information}
\begin{equation}
    C_{\rm cMIMO} = \max_{T_X: {\rm tr}(T_X) \leq P_{\rm T}} \log\det\bigg(I+ \frac{1}{\sigma_z^2}T_h T_X T_h^\dagger\bigg), \label{eqn:H-MIMO_capacity}
\end{equation}
where $T_h: \mathcal{L}^2({\bf L}_t)\to\mathcal{L}^2({\bf L}_r)$ is a bounded linear operator associated with the integral kernel function $h(q,p)$, and $T_X$ is the operator associated with the covariance function $R_X$ of the random field $X$. The symbol $\det(\cdot)$ denotes the Fredholm determinant~\cite{wan2023can} of its argument, and $I$ denotes the identity operator. Note that the signal model in~\eqref{eqn:H-MIMO_capacity} is universal for H-MIMO, XL-MIMO (by allowing the lengths of ${\bf L}_t$ and ${\bf L}_r$ tend to infinity), and discrete MIMO~\eqref{eqn:MIMO_Capacity} (by allowing $R_X$ to contain generalized functions with discontinuities $\delta(\cdot)$).

\subsection{Wavenumber Correlated Channel Model}
Let us focus on the linear Tx/Rx array case. To be specific, ${\bf L}_r = [-L_r/2, L_r/2]$ is the interval where the user receiver operates, $h(q,p)$ is the continuous space channel represented by a Gaussian random field supported on ${\bf L_r}\times {\bf L}_t$, and ${\bf L}_t = [-L_t/2, L_t/2]$ is the interval where the BS transmitter operates. 
For each realization of the random field $h(q,p)$, by applying the Fourier transform, we obtain a random spectrum $\tilde{h}(\beta,\alpha)$ of the original channel field, which is defined as~\cite{nam2014capacity}
\begin{equation}
    \tilde{h}(\beta,\alpha) = \int_{-L_r/2}^{L_r/2} {\rm d}q \int_{-L_t/2}^{L_t/2}{\rm d}p\,h(q,p) e^{-\ri 2\pi q\beta}e^{-\ri 2\pi p\alpha}.  
\end{equation}
Note that for any practical channel ${h}(q,p)$, the channel is intrinsically bandlimited to $[-1,1]$ due to electromagnetic propagation requirements. This means ${\rm supp}(\tilde{h}) \subset [-1,1]^2$. In fact, the spectral points $\alpha, \beta$ are normalized wavenumbers~\cite{pizzo2022fourier,veeravalli2005correlated,poon2005degrees}, with the corresponding physical wavenumber being $\alpha k_c$ and $\beta k_c$, respectively. The most widely adopted random channel model is the separated Kronecker channel model with the correlation function expressed by a product of two white spectra~\cite[Eq. (19)]{bacci2024mmse}.
However, the assumption of white spectra (characterized by correlation functions of $\delta$ singularity at the origin) rarely holds in reality. Since the wavenumber domain channel values $\tilde{h}$ evaluated on adjacent wavenumbers are likely to originate from the same scatterer, they are usually strongly correlated rather than uncorrelated. 
Although the uncorrelated assumption is generally imprecise, this deficiency is somewhat mitigated if a small MIMO aperture is considered. When shifted into advanced MIMO architectures with extremely large-scale apertures or extra-dense antenna placement, the wavenumber domain correlation will be more significant, and thus should be re-emphasized. 

To capture the wavenumber correlation characteristics of the wireless channel, following~\cite{nam2014capacity}, we assume a bandlimited prior to better fit the real-world channel. The channel field $\tilde{h}(\beta, \alpha)$ is compactly supported on $\mathcal{A}_r \times \mathcal{A}_t \subset [-1,1]^2$ and follows a Gaussian bandlimited prior, which is represented by 
\begin{equation}
    \mathbb{E}\big[\tilde{h}(\beta, \alpha)\tilde{h}^*(\beta' ,\alpha')\big] = \frac{\sigma_{\tilde{h}}^2}{\Gamma_t \Gamma_r} \sinc \Big(\frac{\alpha - \alpha'}{\Gamma_t}\Big)\cdot \sinc \Big(\frac{\beta - \beta'}{\Gamma_r}\Big). \label{eqn:SincCorrelation}
\end{equation} 
The presence of the $\rm sinc$ functions in \eqref{eqn:SincCorrelation} imply that the wavenumber domain channel $\tilde{h}$ is bandlimited, with the Tx/Rx-side bandwidths parameterized by bandwidth parameters $1/\Gamma_t$ and $1/\Gamma_r$, respectively. Given the correlation expression~\eqref{eqn:SincCorrelation}, we can apply the Karhunen--Loeve expansion to express the random channel $\tilde{h}$ by an infinite sum of i.i.d.\ randomly weighted basis functions that depends on the spectral supports $\mathcal{A}_{t,r}$, which is given by the PSWF expansion
\begin{equation}
    \tilde{h}(\beta, \alpha) = \sum_{m=0}^{\infty}\sum_{n=0}^{\infty} \tilde{h}_{mn} \psi_{r,m}(\beta)\psi_{t,n}(\alpha), 
\end{equation}
where $\psi_{r,m}$ and $\psi_{t,n}$ are 1D-PSWFs in {\bf Theorem~\ref{thm:Properties_1D_PSWFs}} defined by the pair of compact sets $({\mathcal{A}}_r, \mathcal{B}_{1/\Gamma_r})$ and $({\mathcal{A}}_t, \mathcal{B}_{1/\Gamma_t})$, respectively. The random coefficients $\{\tilde{h}_{mn}\}_{m,n=0}^\infty$ are i.i.d.\ $\mathcal{CN}(0,\sigma_{\tilde{h}}^2)$. Note that, by specifying the prior distribution of $\tilde{h}(\beta, \alpha)$, the distribution of the continuous channel $h(q,p)$ is implicitly defined.

\subsection{Definitions of Continuous Arrays and H-MIMO Systems} \label{sec3_subsec3}
The fundamental difference between a discrete MIMO array and a continuous MIMO array lies in the fact that continuous arrays are capable of generating and measuring a continuously distributed physical quantity (usually the electric field), up to some prescribed precision (noise level). In this subsection, we will introduce a strict mathematical definition of continuous arrays (also called holographic arrays~\cite{sanguinetti2022wavenumber,dardari2020communicating}). 

{\bf Continuous Tx Array.} A continuous Tx array of $n$ elements is a linear map $\phi_t^{(n)}: \mathbb{C}^n \to \mathcal{L}^2({\bf L}_t)$ that satisfies the energy dissipation condition 
\begin{equation}
    \|\phi_t^{(n)}({\bf v})\|_2 \leq 1, \quad\forall\, \|{\bf v}\|_2 = 1.  
\end{equation}
Note that the actual transmitted energy could be strictly less than 1 due to the Ohmic loss inside the antenna structure.  
% Another important concern is that here this $L^2$-norm constraint is imposed on the transmit current density ${\bf j}({\bf s})\,{\rm [A/m^2]}$ of the transmitter spatial region instead of the actual radiated power $P_{\mathsf{rad}}$. In general, the current $\mathcal{L}^2$-norm and $P_{\mathsf{rad}}$ do not vary proportionally~\cite{sanguinetti2022wavenumber}. In fact, we have
% \begin{equation}
%     P_{\mathsf{rad}} \leq Q({\bf L}_t) E_s, 
% \end{equation} 
% where $E_s = \|{\bf j}\|_2^2$ is the squared $\mathcal{L}^2$-norm of the current density ${\bf j}$ defined on the transmitter region ${\bf L}_t$, and the constant factor $Q$ depends on ${\bf L}_t$ through 
% \begin{equation}
%   Q({\bf L}_t) = \frac{k_cZ_0}{4\lambda_c} \sqrt{\iint_{{\bf L}_t\times {\bf L}_t} \sinc^2\left(\frac{2\|{\bf s}_1 - {\bf s}_2\|}{\lambda_c}\right) {\rm d}{\bf s}_1 {\rm d}{\bf s}_2}. 
%\end{equation}
% Generally speaking, constraining the squared $\mathcal{L}^2$ norm $\|\phi\|_2^2$ of the holographic Tx array $\phi$ is interpreted as constraining the upper bound of $P_{\mathsf{rad}}$. However, the gap between $P_{\mathsf{rad}}$ and its upper bound $Q({\bf L}_t) E_s$ is caused by the unradiated energy that has been returned to the antenna structure due to coherent excitation of the antenna array. This portion of energy will be dissipated into heat, instead of being re-radiated into space. Thus, in calculating the power budget, it is more reasonable to impose constraint on the source $\mathcal{L}^2$-norm.  

{\bf Continuous Rx Array.} A continuous Rx array of $n$ elements is a linear map $\phi_r^{(n)}:  \mathcal{L}^2({\bf L}_t) \to \mathbb{C}^n$ that satisfies the energy dissipation condition $\|\phi_r^{(n)}(y)\|_2\leq 1$ for all $y\in \mathcal{L}^2({\bf L}_r)$ such that $\|y\|_2=1$.   

A special case of continuous arrays is the segmented continuous array. This segmented model better approximates real-world antenna arrays that are composed of multiple non-overlapping antenna elements. 

{\bf Segmented Continuous Tx Array.} An $(\delta,{\bf L}_t, n)$-segmented continuous Tx array is a continuous Tx array $\phi_t^{(n)}$ of $n$ elements and antenna spacing $\Delta_t=L_t/n$, which is expressed as  
\begin{equation}
    \begin{aligned}
    [\phi_t^{(n)}({\bf v})](p) &= \frac{1}{\sqrt{\delta}}\sum_{k=1}^{n} [{\bf v}]_k {\rm rect}\left(\frac{2}{\delta}{(p-(k-\frac{n+1}{2})\Delta)}\right), \\
    & \forall p\in {\bf L}_t, 
    \end{aligned} \label{eqn:segmented_holographic_array_def}
\end{equation}
where ${\rm rect}(x) = \mathbbm{1}_{\{|x|\leq 1\}}$ is the rectangular indicator function. Note that $\delta\leq \Delta_t$. 

{\bf Segmented Continuous Rx Array}. An $(\delta, {\bf L}_r, n)$-segmented continuous Rx array is a continuous Rx array $\phi_r^{(n)}$ of $n$ elements and antenna spacing $\Delta_r=L_r/n$, which is expressed as 
\begin{equation}
    [\phi_r^{(n)}(y)]_k = \frac{1}{\sqrt{\delta}} \int_{I_{r,k}} y(q){\rm d}q, \,\forall y\in \mathcal{L}^2({\bf L}_r), \,k\in [n],
    \label{eqn:rsegmented_holographic_array_def}
\end{equation}
where the interval $I_{r,k}$ is defined as $[(k-(n+1)/2)\Delta_r-\delta/2, (k-(n+1)/2)\Delta_r+\delta/2]$, and $\delta\leq \Delta_r$. 
% It is worth noting that, the pre-sum factor $1/\sqrt{\delta}$ has a significant physical meaning of antenna radiation efficiency, which scales inverse-linearly with antenna aperture $\delta$. 

{\bf H-MIMO System.} An $(N_t, N_r)$-holographic MIMO (H-MIMO) transmission system is a discrete MIMO transmission system equipped with a continuous Tx array of $N_t$ elements and a continuous Rx array with $N_r$ elements. 
Although being slightly confusing, we say that H-MIMO systems intrinsically belong to discrete MIMO systems, since every practical H-MIMO system has to operate with a finite number of antenna ports, even though this number can be extremely large.  
Physically speaking, H-MIMO systems are capable of generating and receiving a finite number of electromagnetic field patterns, with each field pattern synthesized with an unlimited spatial resolution. The only difference between H-MIMO and traditional MIMO is that, the antenna spacing of H-MIMO systems can be arbitrarily small, while traditional MIMO systems usually adopt $\lambda_c/2$-spaced antenna arrays.  
Note that according to the definition of continuous arrays, the only two mild requirements are the energy dissipation constraints for Tx and Rx arrays, which are practical physical assumptions. 

In summary, the concept of H-MIMO system is an ideal abstraction of infinitely refined electromagnetic field manipulation and measurement with finite antenna ports. In addition to modern metallic antenna-based RF systems, the definition of H-MIMO system also applies to recently emerging wireless architectures such as atomic receivers.

\subsection{Discrete-Continuous Correspondence}
The EM-based physical transmission models~\eqref{eqn:GreenFunction} are established in the continuous space domain. However, in real-world communication systems the data payload is transmitted and received in a discrete sequential manner. Thus, we need a discrete-continuous correspondence to derive the discrete representation from its continuous counterpart.  

{\bf Discrete-Continuous Correspondence for H-MIMO.} 
Let the continuous-domain channel be $h(q,p)$. Then, for an $(N_t, N_r)$-H-MIMO transmission system, the input-output relationship is given by~\eqref{eqn:DMIMO_channel}, and the discrete channel matrix ${\bf H}\in \mathbb{C}^{N_r\times N_t}$ is induced by the continuous channel operator $T_h$ via
\begin{equation}
    {\bf H} = \phi_r^{(N_r)}\circ T_h \circ \phi_t^{(N_t)}, \label{eqn:discrete_system_induced} 
\end{equation}
where $T_h$ represents the underlying continuous EM channel as in~\eqref{eqn:H-MIMO_capacity}. 
Specifically, for H-MIMO equipped with $(\delta,{\bf L}_t,N_t)$-segmented continuous Tx array and $(\delta,{\bf L}_r, N_r)$-segmented continuous Rx array, the entries of $\bf H$ are calculated as 
\begin{equation}
    \begin{aligned}
        {[{\bf H}]}_{n_r n_t} & = \frac{1}{\delta} \int_{I_{r,n_r}} {\rm d}q \int_{I_{t,n_t}} {\rm d} p \cdot h(q,p), \\
        & n_t \in [N_t], \, n_r \in [N_r], 
    \end{aligned}
\end{equation}
where the intervals are defined as $I_{r, {n_r}} = [(n_r-(N_r+1)/2)\Delta_r-\delta/2, (n_r-(N_r+1)/2)\Delta_r+\delta/2]$, and $I_{t, {n_t}} = [(n_t-(N_t+1)/2)\Delta_t-\delta/2, (n_t-(N_t+1)/2)\Delta_t+\delta/2]$.

\subsection{Signal Model for Discrete Channel Estimation}
Let us consider an $N_r\times N_t$ MIMO channel estimator $\hat{\bf H}$ for the general MIMO signal model~\eqref{eqn:DMIMO_channel} with number of pilots $N_{\rm P}$. At each training time slot $i\in [N_{\rm P}]$, the Tx transmits with a unit-energy precoder ${\bf v}_i\in\mathbb{C}^{N_t\times 1}$, and the Rx receives the MIMO channel output with a unit-energy combiner ${\bf w}_i\in\mathbb{C}^{N_r\times 1}$. Then the pilot transmission model is 
\begin{equation}
    [{\bf y}]_i = \sqrt{P_{\rm T}} {\bf w}_i\H {\bf H} {\bf v}_i + [{\bf z}]_i, 
    \label{eqn:MIMO_CE_formulation}
\end{equation}
where ${\bf y}\in\mathbb{C}^{N_{\rm P}\times 1}$ is the received pilot sequence, ${\bf z}\in\mathbb{C}^{N_{\rm P}\times 1}$ is the AWGN with mean zero and covariance $\sigma_z^2 {\bf I}_{N_{\rm P}}$, and $P_{\rm T}$ is the transmitted power during channel estimation. The goal of a MIMO channel estimator is to recover ${\bf H}$ from the noisy observations ${\bf y}$ by correctly designing $\{{\bf w}_i\}_{i\in[N_{\rm P}]}$ and $\{{\bf v}_i\}_{i\in [N_{\rm P}]}$.

\section{PSWF Bounds for H-MIMO and XL-MIMO} \label{sec4}
In this section, we introduce the discrete-continuous correspondence lemmas that relate a discrete space system to the underlying continuous space system.

\subsection{Discrete-Continuous Correspondence Lemmas}
Modern MIMO transceivers operate in temporally discretized complex baseband sequences with complex-valued matrices of finite dimension. However, in EIT, it is common practice to treat the signals as well as channel responses by continuously indexed random objects (e.g., Gaussian random fields). Although in many works~\cite{pizzo2022spatial,jensen2008capacity} it is intuitively assumed that the continuous performance indicators (e.g., DoF, capacity) should dominate that of the discrete systems, this intuition deserves a theoretical investigation. In fact, it is important to correctly set up the strengths of signals and noises in order to lay a fair foundation for correct discrete-continuous comparison. 

In the following, we establish the exact (non-asymptotic) DoF and capacity comparison results between discrete systems and continuous systems. We recall the matrix and operator $\varepsilon$-DoFs are respectively defined as ${\rm DoF}({\bf H}, \varepsilon) = |\{ i : \sigma_i({\bf H})\ge\varepsilon\}|$ and ${\rm DoF}(T_h, \varepsilon) = |\{ i: \sigma_i(T_h) \geq \varepsilon\}|$ where ${\bf H}$ is a matrix and $T_h$ is an operator. 

\begin{lemma}\label{lemma:DoF_comparison}
    Let the discrete channel ${\bf H}$ be derived from the continuous channel operator $T_h$ via~\eqref{eqn:discrete_system_induced}. Then, for any fixed $\varepsilon \in (0,1)$, the $\varepsilon$-DoF of the discrete system never exceeds that of the underlying continuous system, i.e., 
    \begin{equation}
        {\rm DoF}({\bf H}, \varepsilon) \leq {\rm DoF}(T_h, \varepsilon), \quad \forall\, \varepsilon\in (0,1).
    \end{equation}
    %where the matrix $\varepsilon$-DoF ${\rm DoF}({\bf H}, \varepsilon)$ is defined as $| \{i:\sigma_i({\bf H}) \geq \varepsilon\}|$, and the operator $\varepsilon$-DoF ${\rm DoF}(T_h,\varepsilon)$ is defined as $|\{ i: \sigma_i(T_h) \geq \varepsilon\}|$. 
\end{lemma}

The key ingredients that constitute the proof of {\bf Lemma~\ref{lemma:DoF_comparison}} are the following propositions concerning the spectral dominating properties of an operator and the {\it contracted} versions of the same operator. 
\begin{definition}[Contraction operator] A contraction operator $T: \mathcal{H}_1 \to\mathcal{H}_2$ is a bounded linear operator satisfying $\|Tx\|_2\leq 1$ for all $x\in \mathcal{H}_1$ such that $\|x\|_2\leq 1$. 
\end{definition}
The contraction operator is a mathematical characterization of the fact that all practical systems obey the law of energy dissipation, i.e., the output energy of all real-world passive systems cannot exceed their input energy. Note that contraction matrices are special cases of contraction operators.  

\begin{proposition}[Discrete spectral dominance~\cite{horn2012matrix}] \label{prop:Matrix_SingularValueDominance}
    Let ${\bf P}$ and ${\bf Q}$ be contraction matrices. Then, for any matrix ${\bf A}$, 
    \begin{equation}
        \sigma_i ({\bf PAQ}) \leq \sigma_i ({\bf A}),  \quad \forall i\geq 0. 
    \end{equation}
\end{proposition}

A generalized version of {\bf Proposition~\ref{prop:Matrix_SingularValueDominance}} is the following proposition concerning continuous spectral dominance for contraction operators. 
\begin{proposition}[Continuous spectral dominance]\label{cor:General_spectral_dominance}
    Let $T_P$ and $T_Q$ be contraction operators. Then for any bounded linear operator $T_h$: 
    \begin{equation}
        \sigma_i (T_Q T_h T_P )\leq \sigma_i (T_h), \quad \forall\, i\geq 0.
    \end{equation}
\end{proposition}
\begin{IEEEproof}
    The proof of {\bf Proposition~\ref{cor:General_spectral_dominance}} is detailed in the supplementary material.  
\end{IEEEproof}

From {\bf Proposition~\ref{cor:General_spectral_dominance}}, we can directly derive {\bf Lemma~\ref{lemma:DoF_comparison}} by instantiating $T_P$ to the continuous Tx array response $\phi_t^{(N_t)}$ and $T_Q$ to the continuous Rx array response $\phi_r^{(N_r)}$. 

The spectral dominance result in {\bf Proposition~\ref{cor:General_spectral_dominance}} implies that almost every performance indicator of the discrete systems, including the instantaneous/ergodic rate and instantaneous/average $\varepsilon$-DoF, are dominated by those of the continuous systems. This discrete-continuous comparison result is also shown in~Fig.~\ref{fig:Concepts}. 
The following lemma shows how to bound the capacity of discrete MIMO systems via that of a continuous MIMO system. 

\begin{lemma}[Capacity dominance] \label{lemma:capacity_comparison}
    Let the transmit power be no more than $P_{\rm T}$, and the noise level be $\sigma_z^2$ for both the discrete and the continuous systems. Assume the discrete channel $\bf H$ be derived from the continuous channel operator $T_h$ via~\eqref{eqn:discrete_system_induced}. Then the instantaneous capacity of the discrete system~\eqref{eqn:DMIMO_channel} is upper bounded by that of the underlying continuous system~\eqref{eqn:H-MIMO_signalModel}, i.e., 
    \begin{equation}
        C_{\rm dMIMO} ({\bf H}, P_{\rm T}/\sigma_z^2) \leq C_{\rm cMIMO}(T_h, P_{\rm T}/\sigma_z^2), \quad {\rm a.s.}
        \label{eqn:instantaneous_capacity_dominance}
    \end{equation} 
\end{lemma}
    
\begin{remark}
By applying the expectation operator $\mathbb{E}$ to~\eqref{eqn:instantaneous_capacity_dominance}, we conclude that the ergodic rate of the discrete system does not exceed the underlying continuous system. 
\end{remark}

\subsection{The PSWF Bound for Ergodic Capacity}
Each realization of the random channel $h$ produces a capacity value $C_{\rm cMIMO}$. To obtain some instructive bounds, it is common practice to average over all channel realizations to produce an ergodic capacity. With random channel model~\eqref{eqn:SincCorrelation}, the ergodic capacity limit can be derived by using PSWFs. The following PSWF bound~\cite{nam2014capacity} gives the ergodic capacity upper bound in the case of equal Tx/Rx regions and wavenumber support. 

\begin{lemma}[PSWF bound] \label{lemma:TIT14}
    Let the wavenumber domain correlated channel model be defined as~\eqref{eqn:SincCorrelation} with the same wavenumber support $\mathcal{A}_t=\mathcal{A}_r=\mathcal{A}$, the same bandwidth $1/\Gamma$, and the same Tx/Rx aperture $L$. Then, the ergodic capacity upper bound of the transmission operator $T_h: \mathcal{L}^2({\bf L}_t)\to\mathcal{L}^2({\bf L}_r)$ is determined by the eigenvalues of PSWFs via~\cite[Eq.~(77), (81)]{nam2014capacity}
    \begin{equation}
        \mathbb{E}\left[C_{\rm cMIMO}\left(T_h, \frac{P_{\rm T}}{\sigma_z^2}\right)\right] \leq \sum_{\ell=0}^\infty \log\left(1+\frac{P_{\rm T}}{\sigma_z^2}\gamma_\ell \right),  \label{eqn:ergodic_capacity_bound}
    \end{equation}
    where $\{\gamma_\ell\}_{\ell=0}^\infty$ is the eigenvalue set of the 1D-PSWF defined by the pair of compact sets $(\mathcal{A}, \mathcal{B}_\Omega)$, $\Omega = \min \{\bar{L}, 1/\Gamma\}$, and $\bar{L}=L/\lambda_c$ is the dimensionless normalized array length (also called the electrical length). 
\end{lemma}

\begin{remark}~\label{remark:The_PSWF_bound_is_a_unified_bound}
    By combining~\eqref{eqn:ergodic_capacity_bound} with~\eqref{eqn:instantaneous_capacity_dominance}, the PSWF bound applies to all MIMO systems of normalized aperture $\bar{L}$ with electromagnetic wavenumber correlated channel model~\eqref{eqn:SincCorrelation} and the same SNR constraint. 
\end{remark}

\subsection{Unified Capacity Saturation Effects in H-MIMO and XL-MIMO}
It has been a long-standing debate as to whether H-MIMO can fundamentally outperform massive MIMO (mMIMO)~\cite{sanguinetti2022wavenumber,wei2022multi}. From the capacity perspective, this problem is equivalent to estimating the gap in the discrete-continuous inequality that appear in {\bf Lemma~\ref{lemma:capacity_comparison}}. Fortunately, for any fixed channel operator $T_h$ which is in trace class, one can apply operator SVD to $T_h$ in order to construct the Tx holographic array $\phi_t^{(n)}$ and the Rx holographic array $\phi_r^{(n)}$ with sufficiently large $n$, and thus driving the gap in {\bf Lemma~\ref{lemma:capacity_comparison}} arbitrarily close to 0. Hence, there is no fundamental performance gap between mMIMO and H-MIMO of the same aperture. A similar result has been derived by a Fredholm determinant approach in~\cite{wan2023can}, where a mutual information result was obtained with equal power allocation instead of an ergodic capacity result. 
According to {\bf Remark~\ref{remark:The_PSWF_bound_is_a_unified_bound}}, the H-MIMO capacity will saturate to the PSWF bound as $n\to\infty$. The same capacity bound also applies to mMIMO systems. 

In the XL-MIMO regime, although the array aperture is much larger than the mMIMO case, one may still expect a bounded DoF as the array aperture (the size of channel matrix) tends to infinity. This is because practical transmission systems operate in electromagnetic environments, where the number of eigen-modes supported by the environment cannot be infinity. However, for most of the correlated channel models such as the Kronecker model ${\rm vec}({\bf H}) \sim \mathcal{CN}({\bf 0}, {\bf \Sigma}_t\otimes {\bf \Sigma}_r)$, there will be unbounded DoF and capacity growth as a function of the array aperture. This impractical result arises from the improperly correlated channel model in the XL-MIMO regime. In contrast, with the more practical wavenumber domain correlated channel model~\eqref{eqn:SincCorrelation}, this diverging result is replaced by the PSWF bound~\cite{nam2014capacity} of finite value. 

In summary, the capacity growth of both H-MIMO and XL-MIMO will be finally bounded by the PSWF bound, which we term the {\em unified capacity saturation effect}.

\section{Evaluation of the PSWF Bound} \label{sec5}
In this section, we numerically justify the capacity saturation effect of both H-MIMO and XL-MIMO. Although it is generally difficult to evaluate the bound in {\bf Lemma~\ref{lemma:TIT14}}, there are some techniques that can compute the eigenvalues of PSWFs when the wavenumber domain support $\mathcal{A}$ is well-behaved. In fact, if it is a single interval parameterized by two numbers, i.e., $\mathcal{A} = [a,b]\subset [-1,1]$, then it is possible to numerically evaluate the $(\mathcal{A},\mathcal{B}_\Omega)$-PSWF eigen-system $(\phi_\ell, \gamma_\ell)_{\ell=0}^\infty$ to arbitrarily high precision. Note that the PSWF bandwidth is $\Omega = \min\{\bar{L}, 1/\Gamma\}$. 

\subsection{Solution to the PSWF Integral Equation}
The first step is to transform non-standardized interval $\mathcal{A}=[a,b]$ into the standardized interval $[-1,1]$. This is achieved by the following linear variable substitution $t\to x$: $[a,b]\to[-1,1]$ expressed as 
\begin{equation}
    t = \frac{b(1+x)+a(1-x)}{2}. \label{eqn:linear_rescaling}
\end{equation}
With this substitution, the eigen-equation~\eqref{eqn:PSWF_integral_equation} in terms of eigen-function $\phi$ on $[a,b]$ is converted to 
\begin{equation}
    \begin{aligned}
        \gamma_\ell {\phi}_\ell(x)  = \int_{-1}^{1} & \left(\frac{\Omega(b-a)}{2}\right)\sinc\left(\frac{\Omega(b-a)}{2} (x-y)\right) \\
        & \times {\phi}_\ell (y) {\rm d}y. 
    \end{aligned}\label{eqn:PSWF_eigenequation_1} 
\end{equation}
Let $c = \pi \Omega(b-a)/2$. Then, we can rewrite~\eqref{eqn:PSWF_eigenequation_1} into the standard 1D-PSWF form with eigen-system notation changed to $(\psi_n, \lambda_n)_{n=0}^\infty$. The standard 1D-PSWF eigen-equation then reads
\begin{equation}
    \lambda_n \psi_n(x) = [\mathcal{M}_c\psi_n](x):= \int_{-1}^{1} \frac{\sin (c(x-y))}{\pi(x-y)} \psi_n(y){\rm d}y, 
\end{equation}
where $\mathcal{M}_c$ represents the bandlimited integral operator. With this integral equation formulation, the eigen-system can be solved by the classical Bouwkamp algorithm~\cite{bouwkamp1947spheroidal}. Since this type of PSWF numerical method is generally not familiar to the wireless communications community, we report its main ideas here. These ideas belong to the study of the {\it spectral method} and appears in the mathematical physics~\cite{landau1962prolate3,slepian1964prolate4} literature.

We commence by dealing with the integral operator $\mathcal{M}_c$. This real symmetric integral operator can be split into two simpler finite Fourier transform integral operators~\cite{xiao2001prolate} $\mathcal{M}_c = (c/2\pi) \mathcal{F}_c^\dagger \mathcal{F}_c$, 
where 
\begin{equation}
    [\mathcal{F}_c\psi](x) = \int_{-1}^{1} e^{\ri cxy} \psi(y) {\rm d}y. 
\end{equation} 
We can verify that each eigenvalue $\mu\in\mathbb{C}$ of $\mathcal{F}_c$ corresponds to the eigenvalue $c|\mu|^2/(2\pi)$ of $\mathcal{M}_c$. Furthermore, for each eigenvalue $\lambda>0$ of the positive definite operator $\mathcal{M}_c$, at least one of the four numbers $(\pm, \pm\ri)\cdot\sqrt{2\pi\lambda/c}$ is an eigenvalue of $\mathcal{F}_c$. Thus, the PSWF evaluation problem reduces to evaluating the eigen-system of operator $\mathcal{F}_c$. Fortunately, $\mathcal{F}_c$ commutes\footnote{This is called {\it a lucky accident} in Franceschetti's book~\cite[Sec 2.5.1]{franceschetti2017wave}. However, different from the commuting property $\{\mathcal{M}_c, \mathcal{L}_c\} = 0$ in this book, we alternatively give another commuting property $\{\mathcal{F}_c, \mathcal{L}_c\}=0$, which is easier to deal with.} with the modified Lagrange differential operator $\mathcal{L}_c$~\cite{franceschetti2017wave,xiao2001prolate}, defined as 
\begin{equation}
    \mathcal{L}_c = \frac{\rm d}{{\rm d}x} (1-x^2) \frac{\rm d}{{\rm d}x} - c^2 x^2. \label{eqn:PSWF_differential_operator}
\end{equation}
Since $\mathcal{L}_c$ commutes with $\mathcal{F}_c$, i.e., $[\mathcal{F}_c\mathcal{L}_c\psi](x) = [\mathcal{L}_c\mathcal{F}_c\psi](x)$, these two operators share the same eigenfunctions, which are exactly the PSWFs we want to compute. Note that if $c=0$, the eigen-equation $\mathcal{L}_0 \psi = -\chi \psi$ of operator $\mathcal{L}_0$ turns out to be the Legendre equation. 
Thus, we can write down the eigen-system of $\mathcal{L}_0$ in terms of Legendre polynomials $P_n(x)$. 
Note that Legendre polynomials are orthogonal polynomials with respect to the weight function $w(x) \equiv 1$ on the interval $[-1,1]$, which satisfy the orthogonal property $\int_{-1}^{1}P_m(x) P_n(x) {\rm d}x = \delta_{mn}/(n+1/2)$
and the three-term recursion 
\begin{equation}
    P_{n+1}(x) = \frac{2n+1}{n+1}xP_n(x) - \frac{n}{n+1}P_{n-1}(x), 
    \label{eqn:three_term_recursion}
\end{equation}
where the initial conditions are  $P_0(x) = 1$ and $P_1(x) = x$. However, since Legendre polynomials are not orthonormal, we need to normalize them by defining $\bar{P}_n(x) = \sqrt{n+1/2} P_n(x)$ for analytical convenience. Thus, we can expand the PSWFs $\psi_n$ that satisfy $\mathcal{L}_c \psi_n = -\chi_n \psi_n$~\eqref{eqn:PSWF_differential_operator} with this normalized Legendre polynomials as 
\begin{equation}
    \psi_n(x) = \sum_{k=0}^{\infty} \beta_{nk}(c) \bar{P}_k(x), \quad\forall n\geq 0. \label{eqn:PSWF_expansion_in_Legendre_polynomials} 
\end{equation}
The goal is to determine the expansion coefficients $\beta_{nk}(c)$. since $\{\bar{P}_k\}_{k=0}^\infty$ are eigenfunctions of $\mathcal{L}_0$, we only need to determine the action of the {\em perturbation term} $-c^2x^2$ on the basis $\bar{P}_n$. Thanks to the three-term recursion~\eqref{eqn:three_term_recursion}, we can express $-c^2x^2P_n(x)$ in terms of a linear combination of $P_{n+2}$, $P_{n}$, and $P_{n-2}$~\cite{xiao2001prolate}, 
where $\bar{P}_{-1} = \bar{P}_{-2} = 0$. Thus, under the orthonormal complete basis $\{\bar{P}_n\}_{n=0}^\infty$, the operator $\mathcal{L}_c$ reduces to a penta-diagonal infinite real symmetric matrix ${\bf A}: \ell^2 \to \ell^2$, whose entries are given by 
\begin{equation}
    \begin{aligned}
        {[{\bf{A}}]}_{n, n} &= n(n+1) + \frac{2n(n+1)-1}{(2n+3)(2n-1)}c^2, \\
        [{\bf A}]_{n,n+2} &= \frac{(n+2)(n+1)}{(2n+3)\sqrt{(2n+1)(2n+5)}} c^2, \\
        [{\bf A}]_{n+2, n} &= [{\bf A}]_{n, n+2}, \quad n\geq 0, \\
    \end{aligned} \label{eqn:A_entries}
\end{equation}
and the eigenproblem of differential operator $\mathcal{L}_c$ is finally reduced to a matrix-vector eigenproblem associated with $\bf A$; specifically, $({\bf A} - \chi_n(c) {\bf I}) {\bm \beta}_n(c) = {\bf 0}$, where $[{\bm \beta}_n(c)]_k = \beta_{nk}(c)$. Truncating this infinite-dimensional eigen-equation into a finite-dimensional eigen-problem yields a high-precision numerical solution for the PSWFs $\{\psi_n(x) \}_{n=0}^\infty$ for any $x\in [-1,1]$. The eigenvalues $\{\lambda_n\}_{n=0}^\infty$ can either be computed from applying the operator $\mathcal{F}_c$ to $\psi_n$ numerically, or from the method in~\cite{xiao2001prolate} for improved numerical stability.

The modified Bouwkamp algorithm is summarized in {\bf Algorithm~\ref{alg:Bouwkamp}}. The numerical evaluation operations ``Numerical'' for inner products and operators can be done using  any numerical integral methods, such as the trapezoidal rule (uniform grid nodes with constant grid weights) or Gauss--Legendre quadrature (grid nodes and grid weights determined from Legendre polynomials). 

\begin{algorithm}[!t] 
	\caption{Modified Bouwkamp Algorithm} \label{alg:Bouwkamp}
    \setstretch{1.05}
	\begin{algorithmic}[1]
		\REQUIRE
		Bandwidth parameter $\Omega$; interval $\mathcal{A} = [a,b]$; maximum spectral order $n_{\rm max}$. 
		\ENSURE 
		Approximated PSWFs $\{\phi_\ell(x)\}_{\ell=0}^{n_{\rm max}}$; approximated prolate spheroidal eigenvalues $\{\gamma_\ell\}_{\ell=0}^{n_{\rm max}}$. 
        % \vspace{3pt}
		% \STATE \rule[0.5ex]{1\linewidth}{0.5pt}

        % \STATE {\it \# Stage 1: Computing the prolate spheroidal coefficients}
        \STATE $c\leftarrow \pi\Omega(b-a)/2$. 
        \STATE Construct the normalized Legendre polynomials up to order $n_{\rm max}$ by~\eqref{eqn:three_term_recursion}, and gather the polynomial coefficients in the matrix ${\bf P}\in \mathbb{R}^{(n_{\rm max}+1)\times (n_{\rm max}+1)}$. 
        \STATE Compute [$\bf V, D$] $= {\rm eig}({\bf A})$ of~\eqref{eqn:A_entries} with eigenvalues sorted in descending order.  
        \STATE Let $\beta_{nk}(c) \leftarrow [{\bf V}]_{k,n}$ to determine $\psi_n$ via~\eqref{eqn:PSWF_expansion_in_Legendre_polynomials}. 
        
        \vspace{3pt}
        % \STATE {\it \# Stage 2: Computing the prolate spheroidal eigenvalues}
        \STATE Compute the coefficient matrix ${\bf P}'$ of $\bar{P}_n'(x)$ from ${\bf P}$. 
        \STATE $\mu_0\leftarrow$ Numerical$(\langle \psi_0, F_c\psi_0 \rangle)$.
        \STATE ${y}_{\rm prev}\leftarrow$ Numerical$(\psi_0)$, ${y}_{\rm prev}'\leftarrow$ Numerical$(\psi_0')$. 
        \FOR{$\ell =1,2,\ldots,n_{\rm max}$}
            \STATE ${y}_{\rm cur}\leftarrow$ Numerical$(\psi_\ell)$ by $\beta_{\ell k}(c)$ and $\bf P$. 
            \STATE ${y}_{\rm cur}'\leftarrow$ Numerical$(\psi_\ell')$ by $\beta_{\ell k}(c)$ and ${\bf P}'$. 
            \STATE $a\leftarrow $ Numerical$(\langle {y}_{\rm cur}, {y}_{\rm prev}'\rangle)$. 
            \STATE $b\leftarrow$ Numerical$(\langle {{y}_{\rm cur}'}, {y}_{\rm prev}\rangle)$. 
            \STATE $\mu_\ell\leftarrow \ri \mu_{\ell-1}\cdot\sqrt{|a/b|}$. 
            \STATE ${y}_{\rm prev}\leftarrow {y}_{\rm cur}$, ${y}_{\rm prev}'\leftarrow {y}_{\rm cur}'$. 
        \ENDFOR
        \STATE $\gamma_\ell \leftarrow c|\lambda_\ell|^2/(2\pi),\,\forall \ell\in [n_{\rm max}+1]$. 
        \STATE Get $\phi_\ell$: rescale $\psi_\ell$ from $[-1,1]$ to $[a,b]$ by~\eqref{eqn:linear_rescaling} and applying a re-normalization factor $\sqrt{2/(b-a)}$. 
		\vspace{3pt}
		\RETURN $\{\gamma_\ell\}_{\ell=0}^{n_{\rm max}}$, $\{\phi_\ell\}_{\ell=0}^{n_{\rm max}}$. 
	\end{algorithmic}
\end{algorithm}

\subsection{Numerical Validation of Saturation Effects}
In this subsection, we present numerical results to corroborate the saturation effects predicted by the theoretical analysis on both H-MIMO and XL-MIMO systems. The numerical computation is enabled by {\bf Algorithm~\ref{alg:Bouwkamp}}. 

From the result of {\bf Lemma~\ref{lemma:capacity_comparison}}, we conclude that, with the same transmit energy and noise level, all discrete MIMO systems cannot outperform continuous MIMO systems of the same aperture in terms of their capacities. Then it is of interest to quantify the {\em speed of convergence} of the   capacities of discrete MIMO systems to the capacity of the limiting continuous MIMO system. Generally speaking, performing this quantification requires specifying the type of practical antennas used, e.g., horn antennas, $\lambda_c/4$ antennas, etc. The practical modeling of these antennas is, however, beyond the scope of this paper. To model full aperture efficiency, we assume segmented continuous Tx/Rx arrays with element size $\delta$ that equals to antenna spacing $\Delta_t$, $\Delta_r$, respectively. Following~\cite{poon2006impact}, all the antennas are assumed to be dipoles with no polarization mismatch loss, and the discretized channel matrix is induced by the continuous space channel $h(q,p)$ via
\begin{equation}
    [{\bf H}]_{mn} \approx \min \left\{\frac{\delta}{\lambda_c}, \frac{1}{2}\right\} h(q_m, p_n), 
    \label{eqn:discretized_channel}
\end{equation}
where $\{q_m\}_{m=0}^{N_r-1}\subset {\bf L}_r$ and $\{p_n\}_{n=0}^{N_t-1}\subset {\bf L}_t$ are practical antenna positions with spacings $\Delta_r$ and $\Delta_t$, respectively. Note that due to energy dissipation, the radiated/received power of a single antenna is linearly upper bounded by its aperture $\delta$ in the sub-$\lambda_c/2$ regime. 

\begin{figure}[t]
    \centering
    \includegraphics[width=1\linewidth]{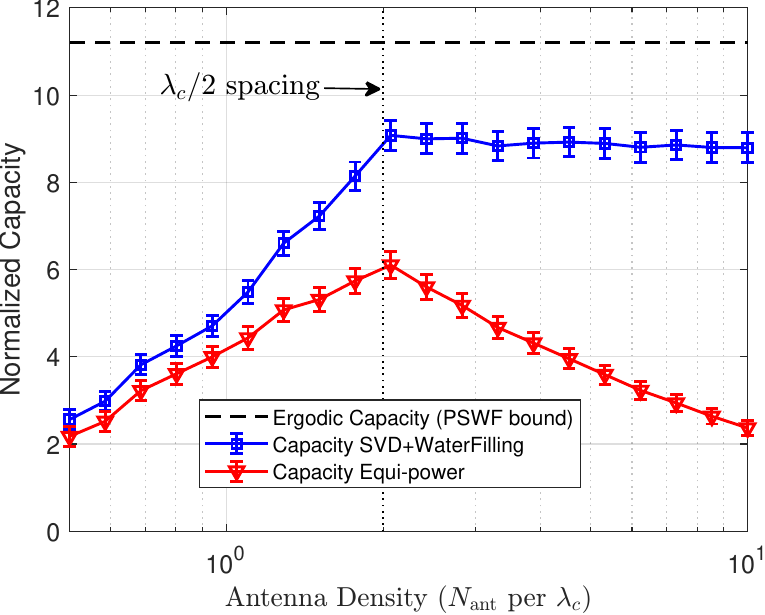}
    \caption{The ergodic capacity saturation phenomenon of H-MIMO with $L_t=L_r=1\,{\rm m}$, $P_{\rm T}/\sigma_z^2 = 10\,{\rm dB}$, and 300 Monte Carlo trials. }
    \label{fig:H-MIMO_capacity}
\end{figure}

In Fig.~\ref{fig:H-MIMO_capacity}, we present the change of ergodic capacity as a function of antenna density $\lambda_c/\delta$ in the H-MIMO regime $\delta\to 0$. The Nyquist spacing case $\Delta_t = \Delta_r = \lambda_c/2$ is given special attention. The capacity of the discretized channel ${\bf H}$ is computed by first performing SVD on $\bf H$ and water-filling to get the optimal ${\bf C_x}$, and then apply~\eqref{eqn:MIMO_Capacity}. The PSWF bound is computed from {\bf Lemma~\ref{lemma:TIT14}} by applying {\bf Algorithm~\ref{alg:Bouwkamp}}. Following~\cite{nam2014capacity}, the wavenumber supports are set to be $\mathcal{A}_t = [-0.85, 0.05]$, $\mathcal{A}_r = [-0.75, 0.15]$, each with total length $0.9$, and the wavenumber domain bandwidth is set to $\Gamma = 0.05$. The equi-power allocation scheme ${\bf C_x} = (P_{\rm T}/N_t) {\bf I}_{N_t}$ is also presented. Note that all the obtained ergodic capacity values are normalized by $\log(1+P_{\rm T}/\sigma_z^2)$. 

Combining Fig.~\ref{fig:H-MIMO_capacity} with the theoretical results, we can make the following conclusions.
\begin{itemize}[leftmargin=10pt]
    \item The PSWF bound~\cite{nam2014capacity} is applicable to discretized MIMO systems through the discrete-continuous correspondence {\bf Lemma~\ref{lemma:DoF_comparison},~\ref{lemma:capacity_comparison}} established in Section~\ref{sec4}. 
    \item With the correct transmitting power and receiving gain normalization~\eqref{eqn:segmented_holographic_array_def}, there is no significant capacity gain by indefinitely densifying MIMO arrays. Even without such a normalization procedure, the DoF will not increase in the H-MIMO regime, as is stated by {\bf Lemma~\ref{lemma:DoF_comparison}}. 
    \item A significant capacity saturation appears, where the transition appears approximately at the Nyquist sampling spacing $\lambda_c/2$. 
\end{itemize} 

In the following, we switch from the H-MIMO regime to the XL-MIMO regime. For XL-MIMO systems, although the antenna spacing is fixed at $\lambda_c/2$, the number of Tx/Rx antennas will grow significantly larger than in traditional mMIMO systems. We adopt the same stochastic channel~\eqref{eqn:SincCorrelation} as in the H-MIMO case with the same parameters. This model captures most of the general behavior of a stochastic channel in the wavenumber domain: the support is finite and relatively small (beamspace sparse), and the channel is slow-varying (bandlimited) as a function of wavenumber. 

\begin{figure}[t]
    \centering
    \includegraphics[width=1\linewidth]{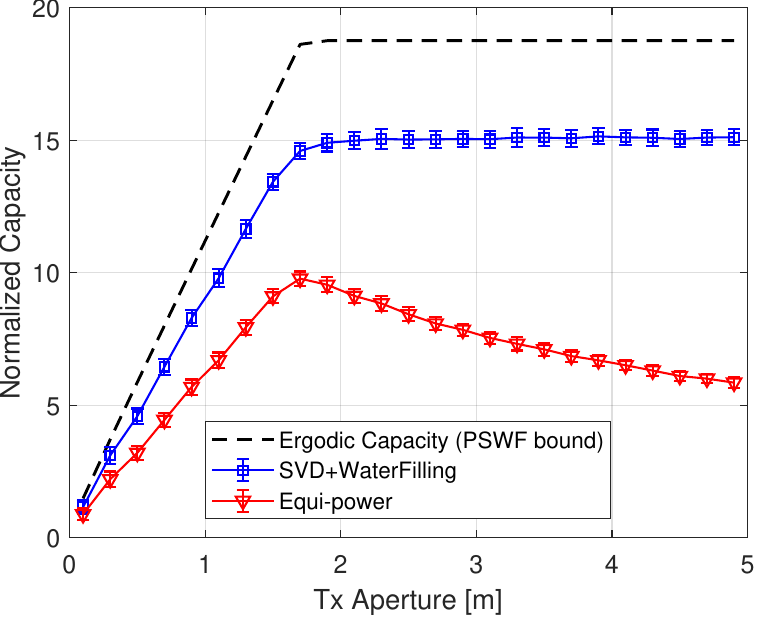}
    \caption{The ergodic capacity saturation phenomenon of XL-MIMO with $\lambda_c/2$ antenna spacing, $P_{\rm T}/\sigma_z^2 = 10\,{\rm dB}$, and 300 Monte Carlo trials. }
    \label{fig:XL-MIMO_capacity}
\end{figure}

Fig.~\ref{fig:XL-MIMO_capacity} shows the dependence of normalized ergodic capacity on the Tx antenna aperture. For symmetry, the Rx antenna aperture is set to equal to the Tx aperture. A clear saturation phenomenon of XL-MIMO is shown in Fig.~\ref{fig:XL-MIMO_capacity}. According to {\bf Lemma~\ref{lemma:TIT14}}, the capacity phase transition from a linear increase to saturation happens when the antenna aperture $L=\lambda_c /\Gamma \approx 1.7\,{\rm m}$, which coincides with the simulation. The PSWF bound correctly predicts the saturation and the location where phase transition occurs. The conclusions are drawn in the following: 
\begin{itemize}[leftmargin=10pt]
    \item The PSWF bound~\cite{nam2014capacity} is applicable to XL-MIMO, which is enabled by the discrete-continuous correspondence {\bf Lemma~\ref{lemma:DoF_comparison},~\ref{lemma:capacity_comparison}} established in Section~\ref{sec4}. 
    \item With the wavenumber bandlimited channel model~\eqref{eqn:SincCorrelation}, there is no significant capacity gain by  enlarging MIMO arrays above the critical aperture $L_{\rm crit} = \lambda_c/\Gamma$ indefinitely. 
\end{itemize}

% the Tx/Rx arrays are more likely to operate within the near-field propagation condition. A widely known effect in the near-field regime is that the steering vectors will change from far-field DFT vectors ${\bf f}$ to near-field vectors ${\bf b}$ featured by quadratic phases. This quadratically phased steering vector ${\bf b}$ is not very sparse in the beamspace, rendering inefficiency of traditional compressed sensing algorithm that rely on far-field codebooks. 

\section{Estimators for the PSWF Channel} \label{sec6}
From the PSWF bound~\eqref{eqn:ergodic_capacity_bound}, it appears that both the capacity of H-MIMO and XL-MIMO are ultimately bounded by some form of continuous ergodic capacity upper bound, which is determined by the underlying EM channel. Thus, generally speaking, continuous EM characteristics pose {\it fundamental limitations} to the performance of wireless MIMO systems. Although it is impossible to exceed the PSWF bound, the mathematical tools of PSWFs can provide us with more {\em opportunities} for more effective and accurate estimation of the underlying channel. 

\subsection{MIMO Channel Estimators}
In this subsection, we operate on the MIMO channel estimation formulation in~\eqref{eqn:MIMO_CE_formulation}. With the given precoder and combiner vectors, we can write the minimum mean squared error (MMSE) estimator as 
\begin{equation}
    \hat{\bf h} = \sqrt{P_{\rm T}} ({\bf C_h}{\bf B}_{\rm MMSE}\H) (P_{\rm T}{\bf B}_{\rm MMSE} {\bf C_h} {\bf B}_{\rm MMSE}\H + \sigma_z^2 {\bf I})^{-1}{\bf y}, 
    \label{eqn:Standard_MMSE_CE}
\end{equation}
where $\hat{\bf h} = {\rm vec}(\hat{\bf H})$, ${\bf h} = {\rm vec}({\bf H})$, ${\bf C_h}\in\mathbb{C}^{N_t N_r \times N_t N_r}$ is the prior covariance matrix of the random vector ${\bf h}$ of length $N_t N_r$, and ${\bf B}_{\rm MMSE}$ is defined via its rows as 
\begin{equation}
    [{\bf B}_{\rm MMSE}]_{i, :} = {\bf v}_i\T \otimes {\bf w}_i\H, \quad i\in [N_{\rm P}]. \label{eqn:B_def}
\end{equation}
The above equation is obtained by vectorizing~\eqref{eqn:MIMO_CE_formulation} and using the matrix vectorization identity ${\rm vec}({\bf ABC}) = ({\bf C}\T \otimes {\bf A}){\rm vec}({\bf B})$. Note that with the full knowledge of ${\bf C_h}$, it is straightforward to design  ${\bf v}_t$ and ${\bf w}_t$ by noticing that the Bayesian MSE $\mathbb{E}[\|{\bf h} - \hat{\bf h}\|^2]$ is a deterministic function of ${\bf v}_t$ and ${\bf w}_t$. However, in general, ${\bf C_h}$ is not fully known to both Tx and Rx. An alternative way is to assume a {\it weaker} prior on ${\bf h}$, e.g., make a sparsity assumption on ${\bf h}$. 

\subsection{Compressed Sensing (CS) Estimators}
By assuming a sparse prior on ${\bf H}$ in the discrete Fourier transform domain representation $\tilde{\bf H}$, i.e., 
\begin{equation}
    {\bf H} = {\bf F}_r \tilde{\bf H} {\bf F}_t\H, 
    \label{eqn:SparseRepresentation}
\end{equation}
the observed signal ${\bf y}$ can be represented by a linear combination of the columns of some dictionary matrix ${\bf B}_{\rm CS}$ with   coefficients given by the entries of $\tilde{\bf h} = {\rm vec}(\tilde{\bf H})$, which is represented by 
\begin{equation}
    [{\bf y}]_i = \sqrt{P_{\rm T}} ({\bf v}_i\T  {\bf F}_t^*)\otimes ({\bf w}_i\H {\bf F}_r) \tilde{\bf h} + [{\bf z}]_i, \quad i\in [N_{\rm P}], 
\end{equation}
where ${\bf F}_t$ and ${\bf F}_r$ are size-$N_t$ and size-$N_r$ DFT matrices, respectively. Using the identity $({\bf AB})\otimes ({\bf CD}) = ({\bf A}\otimes {\bf C})({\bf B}\otimes {\bf D})$, we get ${\bf y} = \sqrt{P_{\rm T}}{\bf B}_{\rm CS}\tilde{\bf h} + {\bf z}$, where ${\bf B}_{\rm CS} = {\bf B}_{\rm MMSE} ({\bf F}_t^* \otimes {\bf F}_r)$. Then, with standard compressed sensing techniques such as approximate message passing (AMP), we can recover the sparse vector $\tilde{\bf h}$ from ${\bf y}$, and thus finally recover ${\bf H}$ via~\eqref{eqn:SparseRepresentation}.

\subsection{Proposed PSWF Channel Estimator}
The performance of MIMO channel estimators is sensitive to the choice of precoders ${\bf v}_i$ and combiners ${\bf w}_i$. Given the prior knowledge on ${\bf h}$, the optimal ${\bf v}_i$ and ${\bf w}_i$ exist theoretically. However, in practice it is almost intractable to accurately estimate the true prior $p({\bf h})$. It is also difficult to efficiently compute the optimal ${\bf v}_i$ and ${\bf w}_i$. Thus, we propose a PSWF channel estimator that heuristically designs the ${\bf v}_i$ and ${\bf w}_i$ by leveraging the structure of the wavenumber bandlimited channel and the properties of the PSWFs in {\bf Theorem~\ref{thm:Properties_1D_PSWFs}}. The proposed PSWF channel estimator (PSWF-CE) operates in a two-step manner. 

{\bf Step 1.} With the channel model~\eqref{eqn:SincCorrelation}, the first step is to estimate the band $\mathcal{A} = [a,b]$ of the wavenumber domain channel $\tilde{h}$. The first step consumes $N_{{\rm P}, 1}$ time slots. Since the center of the wavenumber support, i.e., the center angle of the main scatterer, is slow-varying as as function of time, we assume that the band is centered at 0 with $a=-W/2$ and $b=W/2$, where $W=b-a$ is the wavenumber domain bandwidth. Thus, the band estimation problem is reduced to a bandwidth estimation problem, which is formulated by a maximum {\it a posteriori} (MAP) problem in the form of 
\begin{equation}
    \hat{W}({\bf y}_1) = \argmax_{W\in [0,1]} \log p(W|{\bf y}_1), 
    \label{eqn:BWest}
\end{equation}
where ${\bf y}_1\in\mathbb{C}^{N_{{\rm P}, 1}\times 1}$ is the vector of received signals collected during Step 1 of the PSWF-CE algorithm, and the precoders and combiners are randomly generated with the unit energy constraint. In the MAP formulation~\eqref{eqn:BWest}, we assume a uniform prior on $W$ (i.e., $W\sim {\mathcal{U}}(0,1)$). Thus the log-likelihood function is written (up to a constant) as 
\begin{equation}
    \log p({\bf y}_1 | W) = -{\bf y}_1\H {\bf C}_{W, {\bf y}_1}^{-1} {\bf y}_1 - \log\det {\bf C}_{W, {\bf y}_1} + \mathrm{const},  
\end{equation}
where 
\begin{equation}
    {\bf C}_{W, {\bf y}_1} = P_{\rm T}{\bf B}_1 ({\bf S}_{W, t} \otimes {\bf S}_{W, r}) {\bf B}_1 \H + \sigma_z^2 {\bf I}_{N_{{\rm P},1}}, 
\end{equation}
and ${\bf B}_1$ is similarly defined as ${\bf B}_{\rm MMSE}$ in~\eqref{eqn:B_def} by the randomly generated $\{{\bf w}_i, {\bf v}_i\}_{i=1}^{N_{{\rm P}, 1}}$, and ${\bf S}_{W,(t,r)}$ are size-$N_t$ and size-$N_r$ real symmetric matrices respectively with entries $[{\bf S}_{W, (t,r)}]_{mn}=W\sinc (W(m-n))$. 

{\bf Step 2.} After estimating $W$, we generate ${\bf v}_i$ and ${\bf w}_i$ by computing the length-$N_t$ and length-$N_r$ vectors that are {\it most concentrated} in the spectral interval $[-W/2, W/2]$. This step consumes $N_{{\rm P}, 2}$ time slots. By applying~{\bf Theorem \ref{thm:SimultaneousOrthogonalization}} to the finite-dimensional subspace pair $(\mathcal{H}_1, \mathcal{H}_2)$ defined as 
\begin{equation}
    \mathcal{H}_1 = \{\{a_k\}_{k=0}^{N-1}: a_k \in \mathbb{C} \} = \mathbb{C}^N, 
\end{equation}
and 
\begin{equation}
    \begin{aligned}
    \mathcal{H}_2 = & \bigg\{  \{a_k\}_{k=0}^{N-1}: \sum_{k=0}^{N-1}a_k e^{-\ri 2\pi f k} \equiv 0, \\
    &   \forall f\in [-1/2,1/2]\setminus [-W/2, W/2] \bigg\}, \\
    \end{aligned}
\end{equation}
we know that there exist $N$ unit vectors $\{{\bm \phi}_n\}_{n=1}^{N}$ that can be stacked as  columns of a unitary matrix ${\bf \Phi}\in\mathbb{C}^{N\times N}$ and accompanied by a set of eigenvalues  $\{\tilde{\lambda}_n\}_{n=1}^{N}$. Similar to the PSWF integral equation~\eqref{eqn:PSWF_integral_equation}, we have the matrix equation ${\bf S}_W {\bm \phi}_n = \tilde{\lambda}_n {\bm \phi}_n$,
or 
\begin{equation}
    {\bf S}_W {\bf \Phi} = {\bf \Phi} {\bf \Lambda},  \label{eqn:EigenForm_DPSSs}
\end{equation}
where ${\bf \Lambda}$ is a diagonal matrix with entries $\{\tilde{\lambda}_n\}_{n=1}^{N}$, and ${\bf S}_W\in\mathbb{R}^{N\times N}$ is a symmetric matrix with entries $[{\bf S}_W]_{mn}$ given by $W \sinc (W(m-n))$ for $(m,n)\in [N]^2$. Due to their close connections with PSWFs, the sequence  $\{{\bm \phi}_n\}_{n=1}^{N}$ is known as a  discrete prolate spheroidal sequence (DPSS). From~\eqref{eqn:EigenForm_DPSSs}, we observe that DPSSs can be computed by performing an eigen-decomposition on ${\bf S}_W$. The overall observation-to-estimation procedure of the proposed PSWF-CE is summarized in {\bf Algorithm~\ref{alg:PSWF-CE}}. 

\begin{algorithm}[!t] 
	\caption{PSWF Channel Estimation (PSWF-CE)} \label{alg:PSWF-CE}
    \setstretch{1.05}
	\begin{algorithmic}[1]
		\REQUIRE
		Transmit power $P_{\rm T}$; noise level $\sigma_z^2$; number of pilots during two stages $(N_{{\rm P}, 1}, N_{{\rm P}, 2})$ with total number of pilots $N_{\rm P}$; eigenvalue threshold $\varepsilon$.  
		\ENSURE 
		Estimated channel $\hat{\bf H}$. 
        % \vspace{3pt}
		% \STATE \rule[0.5ex]{1\linewidth}{0.5pt}

        \STATE {\it \# {\bf Step 1}: Bandwidth estimation.}
        \STATE Randomly generate $\{{\bf v}_i, {\bf w}_i\}_{i=1}^{N_{{\rm P}, 1}}$. 
        \STATE Get noisy measurements ${\bf y}_1$ from~\eqref{eqn:MIMO_CE_formulation} by applying the above generated precoders and combiners. 
        \STATE Get bandwidth estimate $\hat{W}$ from~\eqref{eqn:BWest}. 

        \STATE {\it \# {\bf Step 2}: PSWF generation and channel estimation.}
        \STATE Get DPSS $\bf \Phi$ by solving the eigenproblem~\eqref{eqn:EigenForm_DPSSs} of ${\bf S}_{\hat{W}}$.  
        \STATE Randomly choose $\{{\bf v}_i, {\bf w}_i\}_{i=N_{{\rm P}, 1}+1}^{N_{\rm P}}$ from those columns of $\bf \Phi$ whose corresponding eigenvalues are no less than $\varepsilon$. 
        \STATE Get noisy measurements ${\bf y}_2$ from~\eqref{eqn:MIMO_CE_formulation} by the above chosen precoders and combiners. 
        \STATE Construct ${\bf B}_{\rm MMSE}$ from $\{{\bf v}_i, {\bf w}_i\}_{i=1}^{N_{\rm P}}$ via~\eqref{eqn:B_def}. 
        \STATE Compute $\hat{\bf H}$ from~\eqref{eqn:Standard_MMSE_CE} with ${\bf C_h}\leftarrow {\bf I}$. 
		\RETURN $\hat{\bf H}$. 
	\end{algorithmic}
\end{algorithm}

\section{Simulation Results} \label{sec7}
To demonstrate the benefits brought about by PSWFs in EIT, in this section, we present simulation results of the proposed PSWF-based MIMO channel estimation algorithms and compare them to  multiple baselines. 

\subsection{Parameter Settings and Channel Generation}
We numerically simulate the downlink channel estimation of a single-user MIMO communication system operating at $f_c = 3.5\,{\rm GHz}$. Both the Tx and the  Rx are equipped with $\lambda_c/2$-spaced uniform linear arrays. The MIMO channel is generated from the wavenumber domain correlated channel model in Section~\ref{sec3-System-Models-and-Problem-Formulation} with $\mathcal{A}:=\mathcal{A}_t = \mathcal{A}_r = [-0.15, 0.15]$. Thus the true half wavenumber bandwidth is $W/2=0.15$. The wavenumber bandlimited parameter $\Gamma$ is set to $\Gamma=0.05$. The channel generation procedure is mainly based on~\eqref{eqn:SincCorrelation}. Specifically, we discretize the interval $[-1, 1]$ with a uniform grid $\{\theta_k\}_{k=1}^{K}$ comprised of $K=2^{10}$ discrete points. Then, we generate a 2D zero-mean random sequence with correlation $\Gamma^{-2} \sinc (\beta - \beta')\sinc (\alpha - \alpha')$ on the support $\mathcal{A}_r\times \mathcal{A}_t$. This procedure specifies the values $\{\tilde{h}(\theta_m, \theta_n)\}_{m,n=1}^{K}$. Finally, the discrete MIMO channel matrix $\bf H$ is synthesized according to its continuous definition in~\eqref{eqn:SincCorrelation} and the discretization in~\eqref{eqn:discretized_channel}, which is expressed as 
\begin{equation}
    {\bf H} = \sqrt{\Delta_t \Delta_r}\Big(\frac{2}{K}\Big)^2 {\bf D}_r \tilde{\bf H} {\bf D}_t\H, 
\end{equation}
where ${\bf D}_r$ and ${\bf D}_t$ are dictionary matrices with their $k$-th columns corresponding to the array steering vector parameterized by $\theta_k$, and the $(m,n)$-th entry of the matrix $\tilde{\bf H}\in\mathbb{C}^{K\times K}$ is given by the generated random field $\{\tilde{h}(\theta_m, \theta_n)\}_{m,n=0}^{K-1}$. The efficacy of the MIMO channel estimator is quantified by the normalized mean squared error (NMSE), which is expressed as 
\begin{equation}
    {\rm NMSE} = \mathbb{E}\left[\frac{\|{\bf H} - \hat{\bf H}\|_{\rm F}^2}{\|{\bf H}\|_{\rm F}^2}\right]. 
\end{equation}

\begin{table*}[t]
    % \color{blue}
    \centering
    \begin{threeparttable} 
        \caption{Different MIMO Channel Estimators} \label{tab:MIMO_CEs}
        \vspace{-2pt}
        \setstretch{1.1}
        \begin{tabular}{l|c|c|c}
            \hline 
            Channel estimator       & Statistical prior ${\bf C_h}$ & Sparsity prior            & Precoder/combiner design  \\
            \hline
            RandComb MMSE           & Yes                           & Not required              & Random                    \\
            \hline
            RandComb AMP            & Not required                  & Bernoulli-Gaussian        & Random                    \\
            \hline
            BWest PSWF              & Not required                  & Not required              & PSWF                      \\
            \hline
            PSWF w/o StatCSI MMSE   & Not required                  & Interval ${\mathcal{A}}$  & PSWF                      \\
            \hline 
            PSWF w/ StatCSI MMSE    & Yes                           & Interval ${\mathcal{A}}$  & PSWF                      \\ 
            \hline
        \end{tabular}
    \end{threeparttable}
\end{table*}

\subsection{Performance of MIMO Channel Estimators}
In this subsection, we present the simulated performance of the proposed PSWF-CE MIMO channel estimator. We compare the performance of the estimator with standard MMSE baselines and compressed sensing-based approximate message passing (AMP) algorithms. To demonstrate the performance gain brought by PSWF precoders and combiners, we choose the standard MMSE channel estimator and the AMP estimator with random precoders and combiners as the baselines. To show that the proposed PSWF-CE is robust to absence of prior knowledge, we omit both the statistical channel prior and the sparsity prior, yielding the PSWF-CE method with the need of bandwidth estimation~\eqref{eqn:BWest}. For clarity, we compare the properties of the  estimators in {\bf Table~\ref{tab:MIMO_CEs}}. Additionally, we have performed additional simulations on the standardized 3GPP TR 38.901~\cite{CDL} wireless channel model. The results are presented in the {\em supplementary materials}, and the conclusions gleaned from these additional results are the same as those reported here in the main text.

\begin{figure}[t]
    \centering
    \includegraphics[width=1\linewidth]{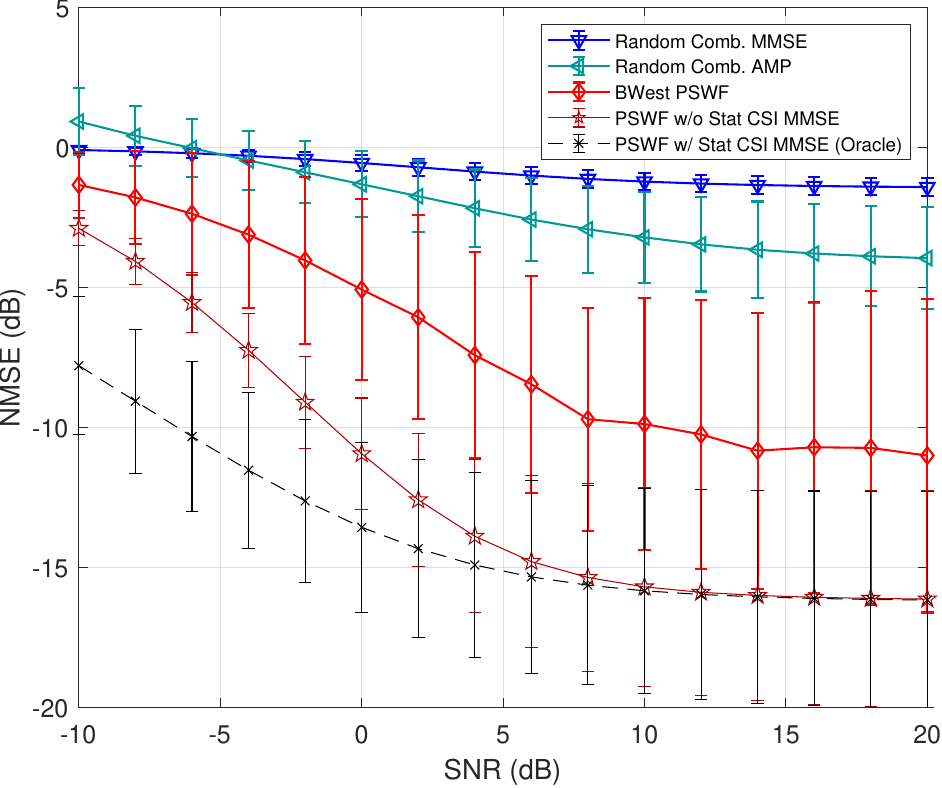}
    \caption{The NMSE performance of MIMO channel estimators with aperture $L_t = L_r = 0.5\,{\rm [m]}$ and $N_{\rm P}=40$. The number of Monte Carlo trials is 100. }
    \label{fig:MIMO_CE_Comb40_wSNR}
\end{figure}

\begin{figure}[t]
    \centering
    \includegraphics[width=1\linewidth]{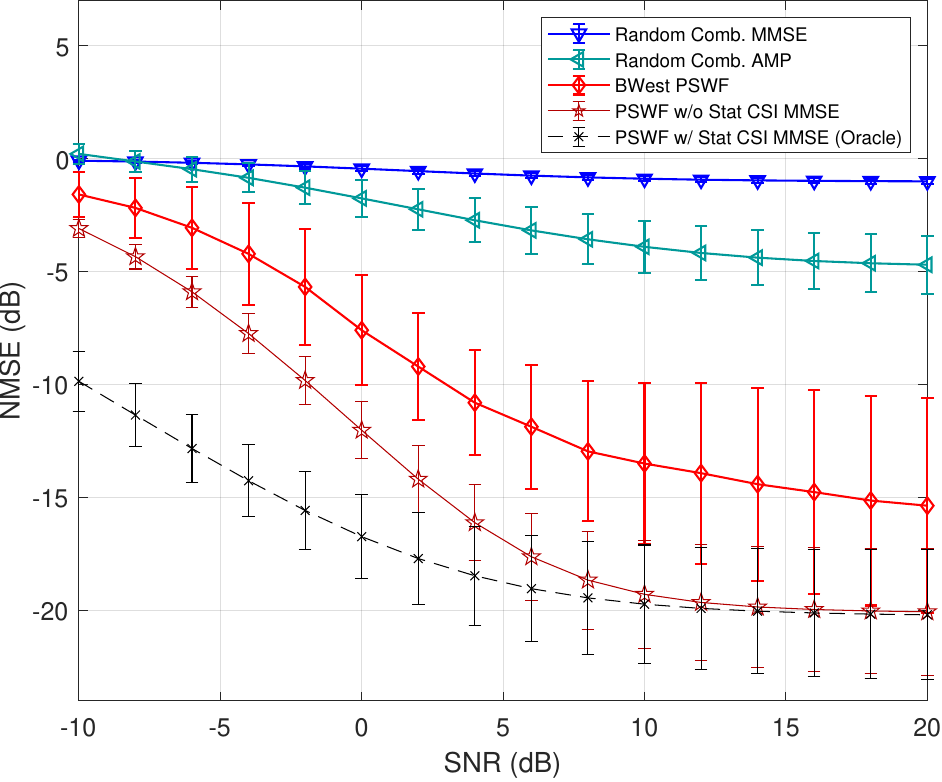}
    \caption{The NMSE performance of MIMO channel estimators with aperture $L_t = L_r = 1.0\,{\rm [m]}$ and $N_{\rm P} = 120$. The number of Monte Carlo trials is 100. }
    \label{fig:MIMO_CE_Comb120_wSNR}
\end{figure}

Fig.~\ref{fig:MIMO_CE_Comb40_wSNR} and Fig.~\ref{fig:MIMO_CE_Comb120_wSNR} display the NMSE performance of MIMO channel estimators with respect to different SNRs $P_{\rm T}/\sigma_z^2$. The error bars indicate one standard deviation above and below the average  NMSE. For Fig.~\ref{fig:MIMO_CE_Comb40_wSNR}, the number of pilots is set to be $N_{\rm P}=40$, compared to the number of unknown channel coefficients $N_tN_r = 144$. Thus, all of the estimators operate on highly compressed measurements with compression ratio $\alpha =40/144\approx 0.278$. The simulation parameters are set in Fig.~\ref{fig:MIMO_CE_Comb120_wSNR} to keep compression ratio the same while increasing the number of Tx/Rx antennas to be $24\times 24$. For the bandwidth estimation (BWest)-PSWF method, the number of pilots for the first step is set to be $N_{{\rm P}, 1} = \lfloor N_{\rm P}/4\rfloor$. From the two figures, it can be concluded that the proposed PSWF estimators outperform traditional AMP and MMSE estimators both in the low- and high-SNR regimes. Among the three PSWF-based estimators, the estimation accuracy increases when there is more prior information available. As SNR increases, the pilot measurements gradually amortizes the influence of the prior information. Thus, the performance of PSWF estimators without prior knowledge of statistical CSI ${\bf C_h}$ eventually approaches the performance of the oracle case in which full statistical CSI is available.   

\begin{figure}[t]
    \centering
    \includegraphics[width=0.96\linewidth]{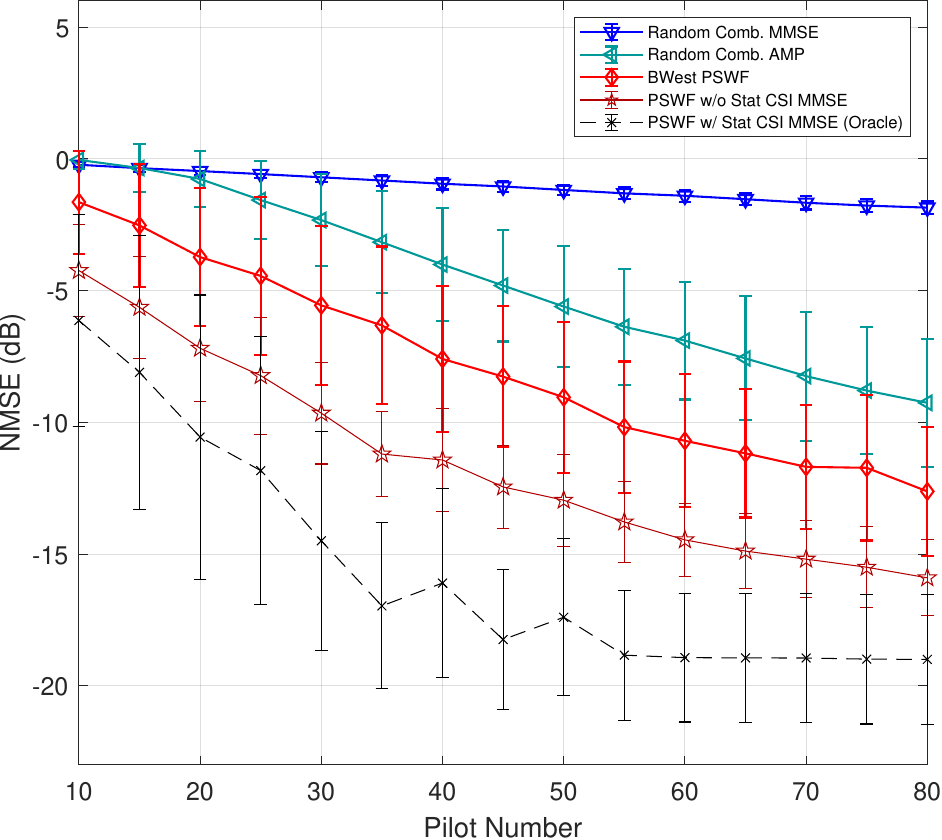}
    \caption{The NMSE performance of MIMO channel estimators with respect to the pilot number $N_{\rm P}$. The Tx/Rx aperture is $L_t = L_r = 0.5\,{\rm [m]}$, ${\rm SNR} = 0\,{\rm dB}$, and the number of Monte Carlo trials is~200. }
    \label{fig:MIMO_CE_wNCombiner_SNR0}
\end{figure}

\begin{figure}[t]
    \centering
    \includegraphics[width=0.96\linewidth]{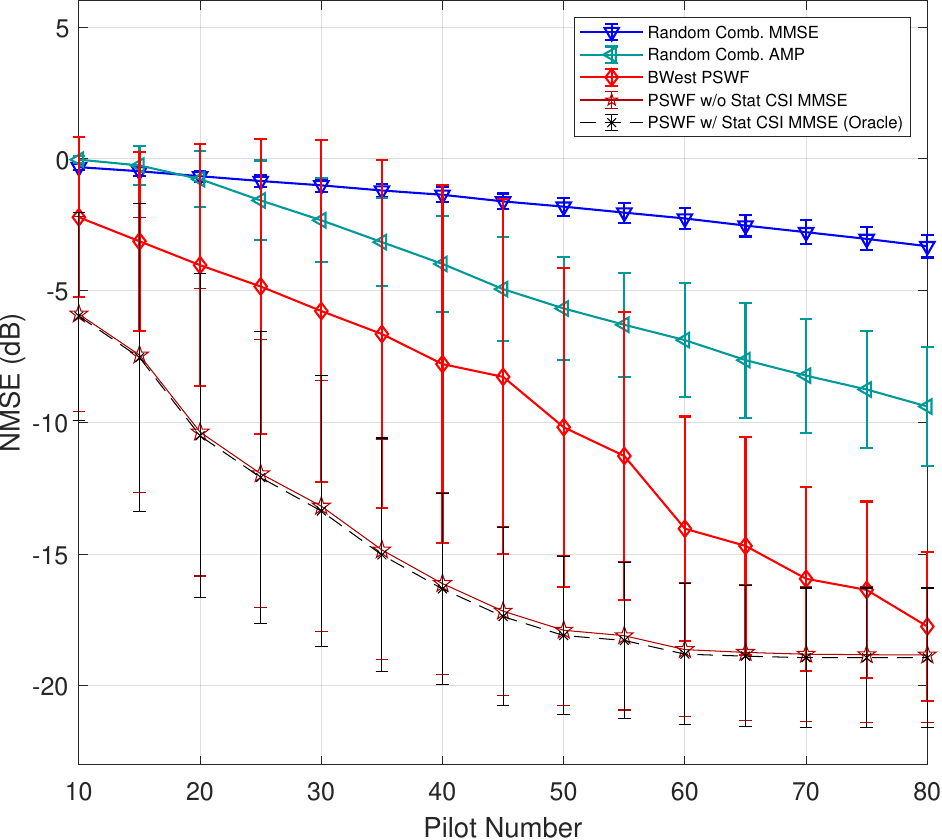}
    \caption{The NMSE performance of MIMO channel estimators with respect to the pilot number $N_{\rm P}$. The Tx/Rx aperture $L_t = L_r = 0.5\,{\rm [m]}$, ${\rm SNR} = 10\,{\rm dB}$, and the number of Monte Carlo trials is 200. }
    \label{fig:MIMO_CE_wNCombiner_SNR10}
\end{figure}

Fig.~\ref{fig:MIMO_CE_wNCombiner_SNR0} and Fig.~\ref{fig:MIMO_CE_wNCombiner_SNR10} display the NMSE performances of the MIMO channel estimators when the  number of pilots $N_{\rm P}$ is varied. Fig.~\ref{fig:MIMO_CE_wNCombiner_SNR0} is plotted with ${\rm SNR}=0\,{\rm dB}$ (low SNR) and Fig.~\ref{fig:MIMO_CE_wNCombiner_SNR10} is plotted with ${\rm SNR}=10\,{\rm dB}$ (moderately high SNR). In the large pilot number regime, the PSWF estimators exhibit a performance saturation around $-20\,{\rm dB}$, which is due mainly to the selection of the threshold of  $\varepsilon = 0.1$ in {\bf Algorithm~\ref{alg:PSWF-CE}}. 
% However, this threshold suffices to enable the proposed PSWF-based estimators to generally outperform both the baselines for a wide range of~$N_{\rm P}$. 

\section{Conclusions} \label{sec8}
In this paper, we  established several fundamental discrete-continuous correspondence lemmas for EIT. These lemmas allow the existing PSWF ergodic capacity bounds to be applicable to more practical discrete space systems. By explicitly evaluating the PSWF eigenvalues via the modified Bouwcamp algorithm, we showed that the ergodic capacity of both H-MIMO and XL-MIMO are subject to some unified upper bound, instead of increasing indefinitely as the antenna aperture and density grows. On the practical side, by applying the concept of continuous PSWF to the problem of discrete pilot design, we showed that the performance of discrete MIMO channel estimators can be readily improved over some baselines methods. Numerical simulations showed that the proposed PSWF-CE method outperforms traditional MMSE and the compressed sensing-based AMP algorithms. 

For future works, the modified Bouwcamp algorithm may be suitably generalized to the case where the PSWF support $\mathcal{A}$ is a union of disjoint intervals, thus enabling the evaluation of ergodic capacity in arbitrary scattering environments. In addition, the wavenumber bandlimited channel model may be extended to the tri-polarized electromagnetic case to increase model accuracy. 

\footnotesize

\bibliographystyle{IEEEtran}
\bibliography{IEEEabrv, bibfile}

% Generated by IEEEtran.bst, version: 1.14 (2015/08/26)
\begin{thebibliography}{10}
\providecommand{\url}[1]{#1}
\csname url@samestyle\endcsname
\providecommand{\newblock}{\relax}
\providecommand{\bibinfo}[2]{#2}
\providecommand{\BIBentrySTDinterwordspacing}{\spaceskip=0pt\relax}
\providecommand{\BIBentryALTinterwordstretchfactor}{4}
\providecommand{\BIBentryALTinterwordspacing}{\spaceskip=\fontdimen2\font plus
\BIBentryALTinterwordstretchfactor\fontdimen3\font minus
  \fontdimen4\font\relax}
\providecommand{\BIBforeignlanguage}[2]{{%
\expandafter\ifx\csname l@#1\endcsname\relax
\typeout{** WARNING: IEEEtran.bst: No hyphenation pattern has been}%
\typeout{** loaded for the language `#1'. Using the pattern for}%
\typeout{** the default language instead.}%
\else
\language=\csname l@#1\endcsname
\fi
#2}}
\providecommand{\BIBdecl}{\relax}
\BIBdecl

\bibitem{migliore2018horse}
M.~D. Migliore, ``Horse (electromagnetics) is more important than horseman
  (information) for wireless transmission,'' \emph{IEEE Trans. Antenna
  Propagat.}, vol.~67, no.~4, pp. 2046--2055, Dec. 2018.

\bibitem{zhu2024electromagnetic}
J.~Zhu, Z.~Wan, L.~Dai, M.~Debbah, and H.~V. Poor, ``Electromagnetic
  information theory: {F}undamentals, modeling, applications, and open
  problems,'' \emph{IEEE Wireless Commun.}, 2024, early access, doi:
  {\color{blue}\href{https://doi-org.libproxy1.nus.edu.sg/10.1109/MWC.019.2200602}{10.1109/MWC.019.2200602}}.

\bibitem{gabor1946theory}
D.~Gabor, ``Theory of communication. {P}art {I}: {T}he analysis of
  information,'' \emph{J. Inst. Electr. Eng.}, vol.~93, no.~26, pp. 429--441,
  Sep. 1946.

\bibitem{bucci1987spatial}
O.~Bucci and G.~Franceschetti, ``On the spatial bandwidth of scattered
  fields,'' \emph{IEEE Trans. Antenna Propagat.}, vol.~35, no.~12, pp.
  1445--1455, Dec. 1987.

\bibitem{bucci1989degrees}
O.~M. Bucci and G.~Franceschetti, ``On the degrees of freedom of scattered
  fields,'' \emph{IEEE Trans. Antenna Propagat.}, vol.~37, no.~7, pp. 918--926,
  Jul. 1989.

\bibitem{migliore2008electromagnetics}
M.~D. Migliore, ``On electromagnetics and information theory,'' \emph{IEEE
  Trans. Antennas Propagat.}, vol.~56, no.~10, pp. 3188--3200, Oct. 2008.

\bibitem{li2022degrees}
H.~Li, ``Degrees of freedom in scattered fields for trade-off in joint
  communications and sensing,'' in \emph{Proc. 2022 IEEE Int'l Conf. Commun.
  (IEEE ICC'22)}.\hskip 1em plus 0.5em minus 0.4em\relax IEEE, May 2022, pp.
  1574--1579.

\bibitem{ding2022degrees}
L.~Ding, E.~G. Str{\"o}m, and J.~Zhang, ``Degrees of freedom in {3D} linear
  large-scale antenna array communications—{A} spatial bandwidth approach,''
  \emph{IEEE J. Sel. Areas Commun.}, vol.~40, no.~10, pp. 2805--2822, Oct.
  2022.

\bibitem{franceschetti2017wave}
M.~Franceschetti, \emph{Wave theory of information}.\hskip 1em plus 0.5em minus
  0.4em\relax Cambridge University Press, 2017.

\bibitem{ruiz2023degrees}
J.~C. Ruiz-Sicilia, M.~Di~Renzo, M.~D. Migliore, M.~Debbah, and H.~V. Poor,
  ``On the degrees of freedom and eigenfunctions of line-of-sight holographic
  {MIMO} communications,'' \emph{\rm arXiv preprint arXiv:2308.08009}, Aug.
  2023.

\bibitem{dardari2020communicating}
D.~Dardari, ``Communicating with large intelligent surfaces: {F}undamental
  limits and models,'' \emph{IEEE J. Sel. Areas Commun.}, vol.~38, no.~11, pp.
  2526--2537, Nov. 2020.

\bibitem{miller2000communicating}
D.~A. Miller, ``Communicating with waves between volumes: {E}valuating
  orthogonal spatial channels and limits on coupling strengths,'' \emph{Applied
  Optics}, vol.~39, no.~11, pp. 1681--1699, Nov. 2000.

\bibitem{poon2005degrees}
A.~S. Poon, R.~W. Brodersen, and D.~N. Tse, ``Degrees of freedom in
  multiple-antenna channels: {A} signal space approach,'' \emph{IEEE Trans.
  Inf. Theory}, vol.~51, no.~2, pp. 523--536, Feb. 2005.

\bibitem{jensen2008capacity}
M.~A. Jensen and J.~W. Wallace, ``Capacity of the continuous-space
  electromagnetic channel,'' \emph{IEEE Trans. Antenna Propagat.}, vol.~56,
  no.~2, pp. 524--531, Feb. 2008.

\bibitem{poon2006impact}
A.~S. Poon, D.~N. Tse, and R.~W. Brodersen, ``Impact of scattering on the
  capacity, diversity, and propagation range of multiple-antenna channels,''
  \emph{IEEE Trans. Inf. Theory}, vol.~52, no.~3, pp. 1087--1100, Mar. 2006.

\bibitem{nam2014capacity}
W.~Nam, D.~Bai, J.~Lee, and I.~Kang, ``On the capacity limit of wireless
  channels under colored scattering,'' \emph{IEEE Trans. Inf. Theory}, vol.~60,
  no.~6, pp. 3529--3543, Apr. 2014.

\bibitem{bjornson2024towards}
E.~Bj{\"o}rnson, C.-B. Chae, R.~W. Heath~Jr, T.~L. Marzetta, A.~Mezghani,
  L.~Sanguinetti, F.~Rusek, M.~R. Castellanos, D.~Jun, and {\"O}.~T. Demir,
  ``Towards 6{G} {MIMO}: {M}assive spatial multiplexing, dense arrays, and
  interplay between electromagnetics and processing,'' \emph{\rm arXiv preprint
  arXiv:2401.02844}, Jan. 2024.

\bibitem{wei2023tri}
L.~Wei, C.~Huang, G.~C. Alexandropoulos, Z.~Yang, J.~Yang, E.~Wei, Z.~Zhang,
  M.~Debbah, and C.~Yuen, ``Tri-polarized holographic {MIMO} surfaces for
  near-field communications: {C}hannel modeling and precoding design,''
  \emph{IEEE Trans. Wireless Commun.}, vol.~22, no.~12, pp. 8828--8842, Apr.
  2023.

\bibitem{sanguinetti2022wavenumber}
L.~Sanguinetti, A.~A. D'Amico, and M.~Debbah, ``Wavenumber-division
  multiplexing in line-of-sight holographic {MIMO} communications,'' \emph{IEEE
  Trans. Wireless Commun.}, Apr. 2022.

\bibitem{lu2023near}
Y.~Lu and L.~Dai, ``Near-field channel estimation in mixed {LoS/NLoS}
  environments for extremely large-scale {MIMO} systems,'' \emph{IEEE Trans.
  Commun.}, vol.~71, no.~6, Jun. 2023.

\bibitem{wong2023fluid}
K.-K. Wong, W.~K. New, X.~Hao, K.-F. Tong, and C.-B. Chae, ``Fluid antenna
  system—{P}art {I}: {P}reliminaries,'' \emph{IEEE Commun. Lett.}, Jun. 2023.

\bibitem{marzetta2019super}
T.~L. Marzetta, ``Super-directive antenna arrays: {F}undamentals and new
  perspectives,'' in \emph{Proc. 53rd Asilomar Conf. Sig. Syst. Comput.}\hskip
  1em plus 0.5em minus 0.4em\relax IEEE, Nov. 2019, pp. 1--4.

\bibitem{liu2021reconfigurable}
Y.~Liu, X.~Liu, X.~Mu, T.~Hou, J.~Xu, M.~Di~Renzo, and N.~Al-Dhahir,
  ``Reconfigurable intelligent surfaces: {P}rinciples and opportunities,''
  \emph{IEEE Commun. Surv. \& Tut.}, vol.~23, no.~3, pp. 1546--1577, Mar. 2021.

\bibitem{wang2022electromagnetic}
T.~Wang, W.~Han, Z.~Zhong, J.~Pang, G.~Zhou, S.~Wang, and Q.~Li,
  ``Electromagnetic-compliant channel modeling and performance evaluation for
  holographic {MIMO},'' in \emph{Proc. IEEE Globecom Workshops}.\hskip 1em plus
  0.5em minus 0.4em\relax IEEE, Dec. 2022, pp. 747--752.

\bibitem{shafi2006polarized}
M.~Shafi, M.~Zhang, A.~L. Moustakas, P.~J. Smith, A.~F. Molisch, F.~Tufvesson,
  and S.~H. Simon, ``Polarized {MIMO} channels in {3-D}: {M}odels, measurements
  and mutual information,'' \emph{IEEE J. Sel. Areas Commun.}, vol.~24, no.~3,
  pp. 514--527, Mar. 2006.

\bibitem{pizzo2020spatially}
A.~Pizzo, T.~L. Marzetta, and L.~Sanguinetti, ``Spatially-stationary model for
  holographic {MIMO} small-scale fading,'' \emph{{IEEE} J. Sel. Areas Commun.},
  vol.~38, no.~9, pp. 1964--1979, Sep. 2020.

\bibitem{pizzo2022spatial}
A.~Pizzo, L.~Sanguinetti, and T.~L. Marzetta, ``Spatial characterization of
  electromagnetic random channels,'' \emph{{IEEE} Open J. Comm. Soc.}, vol.~3,
  pp. 847--866, Apr. 2022.

\bibitem{pizzo2022fourier}
------, ``{F}ourier plane-wave series expansion for holographic {MIMO}
  communications,'' \emph{IEEE Trans. Wireless Commun.}, vol.~21, no.~9, pp.
  6890--6905, Sep. 2022.

\bibitem{mikki2023shannon}
S.~Mikki, ``The {S}hannon information capacity of an arbitrary radiating
  surface: {A}n electromagnetic approach,'' \emph{IEEE Transactions on Antennas
  and Propagation}, vol.~71, no.~3, pp. 2556--2570, Mar. 2023.

\bibitem{li2023electromagnetic}
R.~Li, D.~Li, J.~Ma, Z.~Feng, L.~Zhang, S.~Tan, E.~Wei, H.~Chen, and E.-P. Li,
  ``An electromagnetic information theory based model for efficient
  characterization of {MIMO} systems in complex space,'' \emph{IEEE Trans.
  Antenna Propagat.}, vol.~71, no.~4, pp. 3497--3508, Jan. 2023.

\bibitem{castellanos2023electromagnetic}
M.~R. Castellanos and R.~W. Heath~Jr, ``Electromagnetic manifold
  characterization of antenna arrays,'' \emph{\rm arXiv preprint
  arXiv:2311.04835}, Nov. 2023.

\bibitem{wei2022multi}
L.~Wei, C.~Huang, G.~C. Alexandropoulos, E.~Wei, Z.~Zhang, M.~Debbah, and
  C.~Yuen, ``Multi-user holographic {MIMO} surfaces: {C}hannel modeling and
  spectral efficiency analysis,'' \emph{IEEE J. Sel. Top. Sig. Process.},
  vol.~16, no.~5, pp. 1112--1124, May 2022.

\bibitem{ghermezcheshmeh2023parametric}
M.~Ghermezcheshmeh and N.~Zlatanov, ``Parametric channel estimation for {LoS}
  dominated holographic massive {MIMO} systems,'' \emph{IEEE Access}, vol.~11,
  no.~5, pp. 44\,711--44\,724, May 2023.

\bibitem{gruber2008new}
F.~K. Gruber and E.~A. Marengo, ``New aspects of electromagnetic information
  theory for wireless and antenna systems,'' \emph{IEEE Trans. Antenna
  Propagat.}, vol.~56, no.~11, pp. 3470--3484, Nov. 2008.

\bibitem{kong1975theory}
J.~A. Kong, ``Theory of electromagnetic waves,'' \emph{New York}, 1975.

\bibitem{slepian1964prolate4}
D.~Slepian, ``Prolate spheroidal wave functions, {F}ourier analysis and
  uncertainty-{IV}: {E}xtensions to many dimensions; {G}eneralized prolate
  spheroidal functions,'' \emph{Bell System Technical J.}, vol.~43, no.~6, pp.
  3009--3057, Jun. 1964.

\bibitem{ihara1993information}
S.~Ihara, \emph{Information theory for continuous systems}.\hskip 1em plus
  0.5em minus 0.4em\relax World Scientific, 1993.

\bibitem{wan2023can}
Z.~Wan, J.~Zhu, and L.~Dai, ``Can continuous aperture {MIMO} obtain more mutual
  information than discrete {MIMO}?'' \emph{IEEE Commun. Lett.}, Nov. 2023.

\bibitem{veeravalli2005correlated}
V.~V. Veeravalli, Y.~Liang, and A.~M. Sayeed, ``Correlated {MIMO} wireless
  channels: {C}apacity, optimal signaling, and asymptotics,'' \emph{IEEE Trans.
  Inf. Theory}, vol.~51, no.~6, pp. 2058--2072, Jun. 2005.

\bibitem{bacci2024mmse}
G.~Bacci, A.~A. D'Amico, and L.~Sanguinetti, ``{MMSE} channel estimation in
  large-scale {MIMO}: {I}mproved robustness with reduced complexity,''
  \emph{\rm arXiv preprint arXiv:2404.03279}, Apr. 2024.

\bibitem{horn2012matrix}
R.~A. Horn and C.~R. Johnson, \emph{Matrix analysis}.\hskip 1em plus 0.5em
  minus 0.4em\relax Cambridge university press, 2012.

\bibitem{bouwkamp1947spheroidal}
C.~Bouwkamp, ``On spheroidal wave functions of order zero,'' \emph{J. Math.
  Physics}, vol.~26, no. 1-4, pp. 79--92, Apr. 1947.

\bibitem{landau1962prolate3}
H.~J. Landau and H.~O. Pollak, ``Prolate spheroidal wave functions, {F}ourier
  analysis and uncertainty-{III}: {T}he dimension of the space of essentially
  time-and band-limited signals,'' \emph{Bell System Technical J.}, vol.~41,
  no.~4, pp. 1295--1336, Apr. 1962.

\bibitem{xiao2001prolate}
H.~Xiao, V.~Rokhlin, and N.~Yarvin, ``Prolate spheroidal wavefunctions,
  quadrature and interpolation,'' \emph{Inverse Prob.}, vol.~17, no.~4, pp.
  805--838, Apr. 2001.

\bibitem{CDL}
\emph{Study on channel model for frequencies from 0.5 to 100 {G}Hz ({R}elease
  16) v16.1.0}, 3GPP TR 38.901 TSG RAN; NR, Dec. 2019.

\end{thebibliography}

\end{document}